\documentclass[9pt,twoside,web]{ieeecolor}
\pdfoutput=1
\usepackage{generic}
\usepackage{cite}
\usepackage{amsmath,amssymb,amsfonts}
\usepackage{algorithmic}
\usepackage{graphicx}
\usepackage{threeparttable}
\usepackage{booktabs}
\usepackage[backref]{hyperref}
\usepackage{textcomp}

\usepackage{subcaption}
\usepackage{wrapfig}

\usepackage{framed,multirow}
\usepackage{latexsym}
\usepackage{url}
\usepackage{xcolor}
\usepackage{soul}
\usepackage{lineno}
\usepackage{makeidx} 
\usepackage{caption}
\usepackage{color}
\usepackage{bm}
\usepackage{comment}

\usepackage[section]{placeins}

\usepackage{ltxtable, filecontents}

\begin{document}
\title{Deep Learning in Breast Cancer Imaging: A Decade of Progress and Future Directions}
\author{
Luyang Luo,~\IEEEmembership{Member,~IEEE,}
Xi Wang,
Yi Lin,
Xiaoqi Ma,
Andong Tan,
Ronald Chan,
Varut Vardhanabhuti,
Winnie CW Chu,
Kwang-Ting Cheng,~\IEEEmembership{Fellow,~IEEE,}
Hao Chen*,~\IEEEmembership{Senior~Member,~IEEE}
\thanks{This work was supported by National Natural Science Foundation of China (No. 62202403), Shenzhen Science and Technology Innovation Committee (Project No. SGDX20210823103201011), Hong Kong Innovation and Technology Fund (No. PRP/034/22FX) and in part by the Project of Hetao Shenzhen-Hong Kong Science and Technology Innovation Cooperation Zone (HZQB-KCZYB-2020083).}
\thanks{Luyang Luo, Yi Lin, Xiaoqi Ma, Andong Tan, Kwang-Ting Cheng, and Hao Chen are with the Department of Computer Science and Engineering, The Hong Kong University of Science and Technology.}
\thanks{Xi Wang is with the Department of Radiation Oncology, Stanford University School of Medicine.}
\thanks{Ronald Chan is with the Department of Anatomical and Cellular Pathology, The Chinese University of Hong Kong, Hong Kong.}
\thanks{Varut Vardhanabhuti is with the Department of Diagnostic Radiology, Li Ka Shing Faculty of Medicine, The University of Hong Kong.}
\thanks{Winnie CW Chu is with the Department of Imaging and Interventional Radiology, The Chinese University of Hong Kong.}
\thanks{Kwang-Ting Cheng is also with the Department of Electronic and Computer Engineering, The Hong Kong University of Science and Technology.}
\thanks{Hao Chen is also with the Department of Chemical and Biological Engineering, The Hong Kong University of Science and Technology, Hong Kong, China; and HKUST Shenzhen-Hong Kong Collaborative Innovation Research Institute, Futian, Shenzhen, China.}
\thanks{* Hao Chen is the corresponding author (email:jhc@cse.ust.hk).}
}

\maketitle

\begin{abstract}
Breast cancer has reached the highest incidence rate worldwide among all malignancies since 2020.
Breast imaging plays a significant role in early diagnosis and intervention to improve the outcome of breast cancer patients.
In the past decade, deep learning has shown remarkable progress in breast cancer imaging analysis, holding great promise in interpreting the rich information and complex context of breast imaging modalities.
Considering the rapid improvement in deep learning technology and the increasing severity of breast cancer, it is critical to summarize past progress and identify future challenges to be addressed.
This paper provides an extensive review of deep learning-based breast cancer imaging research, covering studies on mammogram, ultrasound, magnetic resonance imaging, and digital pathology images over the past decade.
The major deep learning methods and applications on imaging-based screening, diagnosis, treatment response prediction, and prognosis are elaborated and discussed.
Drawn from the findings of this survey, we present a comprehensive discussion of the challenges and potential avenues for future research in deep learning-based breast cancer imaging.
\end{abstract}


\section{Introduction}

Breast cancer has become the malignancy with the highest incidence rate worldwide with estimated 2.3 million new cases in 2020 \cite{harbeck2017breast}.
Although the mortality rate has steadily decreased since 1989 \cite{giaquinto2022breast}, breast cancer remains the fifth leading cause of cancer mortality globally and the primary cause of cancer mortality in women, with an increasing incidence rate in most of the past four decades and an estimated 685,000 deaths in 2020 \cite{giaquinto2022breast,sung2021global}.

Breast cancer can be categorized into invasive cancer and in situ cancer according to whether \textcolor{black}{it} spreads out or not, and invasive cancer is further divided into four stages (i.e., I, II, III, or IV) based on the spreading severity \cite{amin2017eighth}.
Recent statistics by the American Cancer Society showed that breast cancer survival varies significantly by stage at diagnosis. The 5-year survival rates of USA patients diagnosed during 2012-2018 were $>$99\% for stage I, 93\% for stage II, 75\% for stage III, and 29\% for stage IV \cite{giaquinto2022breast}.
Early detection and efficient systemic therapies are essential in reducing the mortality rate of breast cancer \cite{harbeck2017breast}.
Breast imaging, including mammography, ultrasonography, magnetic resonance imaging, and pathology imaging, has played a crucial role in providing both macroscopic and microscopic investigation of breast cancer to guide treatment decisions. 

\begin{figure*}[t]
  \centering
    \includegraphics[width=\textwidth]{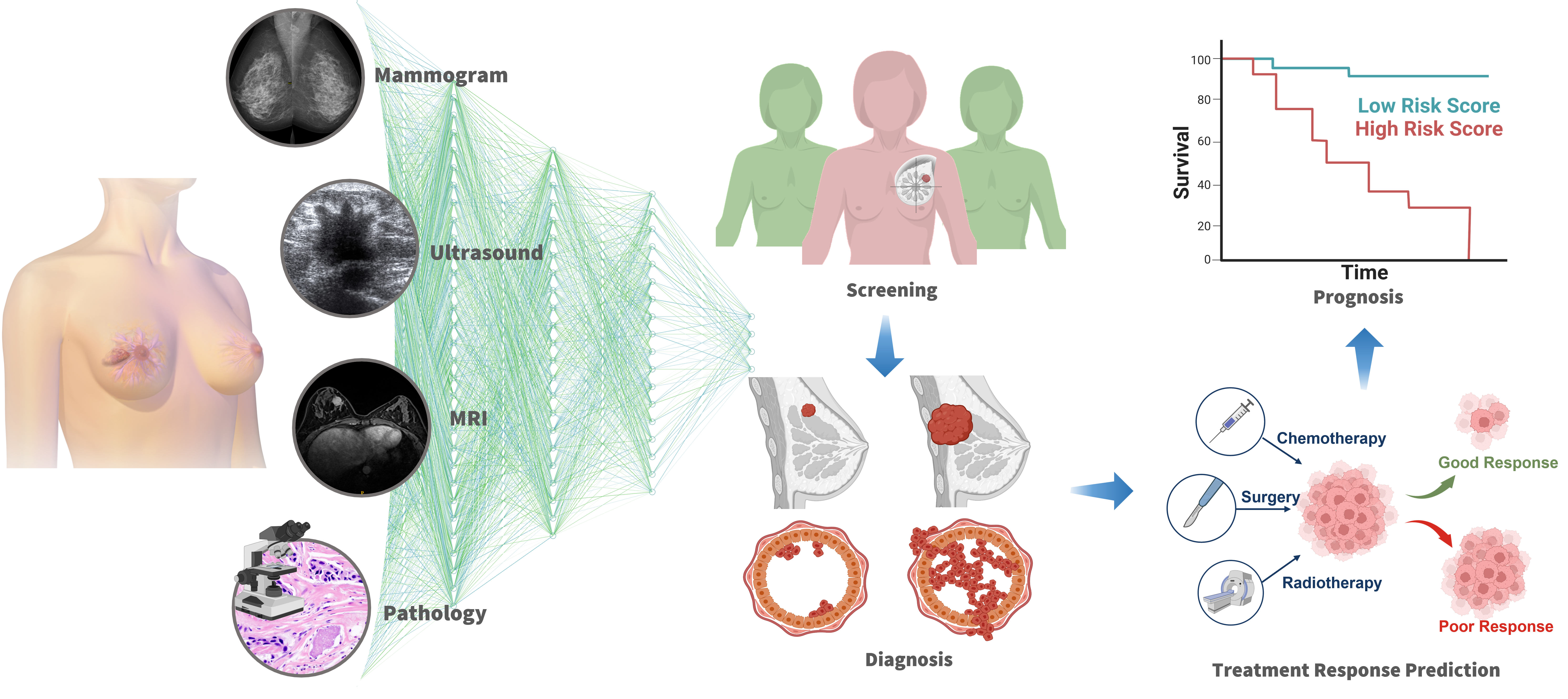}
  \caption{Overview of deep learning in breast cancer imaging. Typical imaging techniques include mammogram, ultrasound, magnetic resonance imaging (MRI), and pathology images. Deep learning is often used for screening, diagnosis, treatment response prediction, and prognosis.}
  \label{fig:intro}
\end{figure*}

Mammography, first performed in 1913, has been proven able to reduce breast cancer mortality rates after long-term follow-up \cite{nystrom2002long}.
Mammography uses low-energy X-rays to examine the breast, often by projecting the tissues into a 2D image.
Organized population-based mammography screening for women is recommended by The World Health Organization \cite{world2014position}, which can provide early diagnosis and improve prognosis for potential patients \cite{marmot2013benefits}.
Apart from screening, mammograms are also used to diagnose abnormalities such as masses, calcifications, architecture distortions, or area asymmetries.
Due to the superposition nature of 2D mammograms, different views of a breast could be needed for richer information.
Standard mammography views are the craniocaudal (CC) view and the mediolateral oblique (MLO) view for both breasts, which are taken directly from the above and from an angled side of the breast, respectively.
Extra views might also be generated depending on \textcolor{black}{practical} needs.
Digital breast tomosynthesis (DBT), also regarded as 3D mammography, has been introduced to provide more spatial context for detailed examination of the breasts and is emerging as the standard of breast imaging care \cite{chong2019digital}.

Ultrasound imaging (sonography) uses high-frequency sound waves to view inside the body without any ionizing radiation. 
Since the early attempts in describing the acoustic characteristics of the breast tumors \cite{wild1951use}, ultrasonic imaging has undergone a series of transformations, both in instrument design and in clinical applications. 
Over the past few decades, the quality of ultrasound images has been largely improved by advances in transducer design, electronics, computers, and signals.
Sonography thus has become a major mode of imaging for the diagnosis of breast cancer in clinical practice~\cite{sehgal2006review}. 
Currently, breast ultrasound is widely used to distinguish cysts and solid nodules with a high specificity~\cite{hooley2013breast} and classify solid masses as benign or malignant when combined with mammography~\cite{ kapur2004combination}. 
It has also shown usefulness in screening and detecting early-stage breast cancers \cite{berg2008combined}, and is recommended for Asian women with dense breasts \cite{leong2012supplementary}.
Due to its ease of use and real-time imaging capability, breast ultrasound becomes popular in guiding breast biopsies and other interventional procedures.
B-mode is the most common form of ultrasonic imaging for the breasts. Compound imaging and harmonic imaging are also increasingly applied to visualize breast lesions and reduce image artifacts. 
Moreover, there is growing interest in applying Colored Doppler ultrasound and contrast agents for measuring tumor blood flow and imaging tumor vascularity~\cite{sehgal2000quantitative}.

Breast magnetic resonance imaging (MRI) \cite{mann2019breast} takes advantage of radio waves and magnetic fields to generate more detailed information, which is often a 3D picture of the inside of the breasts.
Since the invention of MRI in 1971, multiple clinical assessments have witnessed the versatility and effectiveness of breast MRI.
Breast MRI has the highest sensitivity among radiological imaging techniques for breast cancer detection \cite{mann2019breast}, and it is widely used as an auxiliary tool for breast-related lesion diagnosis and prognosis.
Nowadays, MRI examinations are becoming the main scanning modalities for monitoring the cycle treatment response and recurrence, offering more details of the breasts without introducing ionizing radiations.
Considering that the breasts anatomy contains \textcolor{black}{different types of tissues}, fat suppression technique \cite{kalovidouri2017fat} has been introduced to suppress the signal from adipose tissue as an auxiliary step. 
To provide different visible foci, multiple types of sequences could be generated \cite{bernstein2004handbook}, such as T1-weighted, T2-weighted, and Diffusion-weighted MRI.
Moreover, Dynamic Contrast Enhanced (DCE)-MRI has become the main clinical and research sequence, which could provide additional information by observing the T1 changes over multiple phases after injection of the contrast agent \cite{2003Correlation}.
Abbreviated breast MRI which uses single early phase DCE has been introduced as a shortened examination approach for screening breast cancers \cite{gao2020abbreviated}.

Breast pathology provides a microscopic investigation for cancers in an invasive way.
In clinical practice, microscopic analysis by pathology imaging is also regarded as the gold standard for the final determination of breast cancer.
A sample of the patient's breast tissue would be taken by pathologists and placed onto a microscope slide. 
Then, specific stains and dyes are used to identify cancer cells and confirm the presence of chemical receptors.
The most common stain for breast tissue specimens is the hematoxylin-eosin stain (H\&E stain)~\cite{duregon2014comparative}, which has been used for more than a century and is still the standard process for histopathological diagnosis~\cite{klaiman2018enabling}.
Moreover, auxiliary techniques are often required to complete the diagnosis, such as immunohistochemistry (IHC)~\cite{liu2014application} and in situ hybridization (ISH)~\cite{hanna2014her2}. 
In the routine clinical pathology, the predictive and prognostic biomarkers estrogen receptor $\alpha$ (ER), progesterone receptor (PgR), human epidermal growth factor receptor 2 (HER2), and the proliferation-associated nuclear protein Ki67 are analyzed by IHC~\cite{coates2015tailoring}. 
HER2 gene amplification can be further verified by ISH analysis~\cite{hanna2014her2}.

The breast imaging-reporting and data system (BI-RADS) was proposed to categorize the overall assessment of the radiological imaging findings \cite{magny2022breast}:
BI-RADS 0 refers to an incomplete examination;
BI-RADS 1 refers to negative findings;
BI-RADS 2 refers to benign findings;
BI-RADS 3 refers to likely benign findings with $<$2\% chance of malignancy;
BI-RADS 4 has three sub-categories, 4a, 4b, and 4c, which refer to suspicious findings with 2\%-10\%, 10\%-50\%, and 50\%-95\% likelihood of malignancy, respectively;
BI-RADS 5 refers to suspicious findings with $>$95\% likelihood of malignancy;
and BI-RADS 6 refers to pathology-proven malignancy.
The radiological findings can only be used as a reference for suspicion of malignancy.
Usually, patients with BI-RADS 4 or above would be recommended for a biopsy examination to determine the status of malignancy in a microscopic view.

The description of breast cancer requires interpretation of the complex and rich clinical information provided by breast imaging from the macroscopic level to the microscopic level.
With the fast increase in medical data scale and the development of imaging technology, analyzing large-scale high-dimensional breast images with artificial intelligence (AI) holds great promise in improving the accuracy and efficiency of clinical procedures.
Current AI is typically represented by deep learning (DL), which has made remarkable achievements over the past decade and has been widely adopted in various fields such as image or speech recognition \cite{lecun2015deep}.
Compared with conventional computer-aided diagnosis techniques that rely on hand-engineered features, deep learning models show great efficacy in extracting representations from high-dimensional data (e.g., images), and the performance of deep models is often better with more training data.
\textcolor{black}{Thus far, deep learning has also been widely studied for analyzing medical images \cite{litjens2017survey} and demonstrated high performance in various fields \cite{topol2019high}.}
With the convergence of AI and human performance, deep learning nowadays is also reshaping cancer research and personalized clinical care.

\begin{figure}[t]
\centering
 \includegraphics[width=0.48\textwidth]{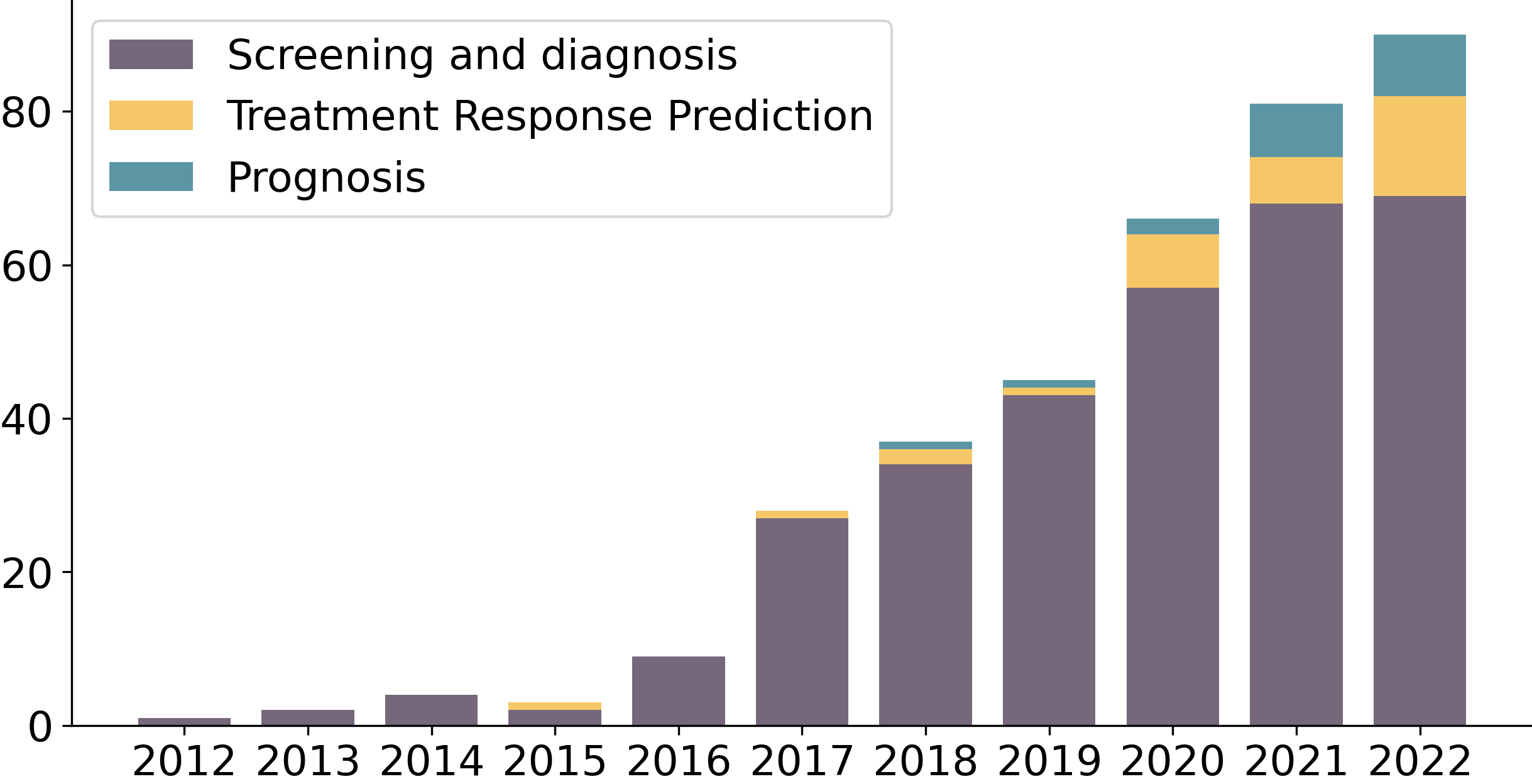}
  \caption{Number of representative papers on deep learning for breast cancer imaging published from 2012 to 2022.}
  \label{fig:Trend}
\end{figure}

As shown in Fig. \ref{fig:Trend}, deep learning-based breast imaging has a prosperous development in the past decade.
However, an extensive survey on deep learning-based breast cancer analysis is yet absent to narrate the progress in various imaging modalities over the past decade.
Therefore, the main goal of this paper is to review the development of deep learning in breast cancer imaging, identify the challenges yet to be addressed in this field, and highlight potential solutions to these challenges.
Specifically, this survey includes applications from screening, diagnosis, and treatment response prediction to prognosis, covering imaging modalities from mammography, ultrasound, and MRI, to pathology images.
Compared with previous surveys that focus on one or two specific modalities \cite{bai2021applying,duggento2021deep,hamidinekoo2018deep,geras2019artificial,sechopoulos2021artificial}, this work provides a more comprehensive summary of the advances in this field.
In total 366 papers from 2012 to 2022 were surveyed, covering a wide variety of applications of deep learning in breast cancer imaging.

The remainder of this work is structured as follows: In Section \ref{Deep Learning Methods for Breast Cancer Analysis}, we introduce the major deep learning techniques used in breast cancer image analysis.
In Section \ref{Applications in Breast Cancer}, we elaborate in detail on the applications of deep learning in breast cancer image analysis in four aspects: screening, diagnosis, treatment response prediction, and prognosis. 
In Section \ref{Challenges and Future Directions}, we discuss the major challenges facing the field and highlight the future perspectives that hold promise in advancing the field. 
Finally, we conclude this survey in Section \ref{conclusion}.
\textcolor{black}{We also summarize publicly available datasets and provide a more detailed table of surveyed papers in the Supplementary Materials for interested readers.}

\section{Deep Learning Methods for Breast Cancer Analysis}
\label{Deep Learning Methods for Breast Cancer Analysis}

\begin{figure*}[t]
\centering
\begin{subfigure}{\linewidth}
  \centering
  \includegraphics[width=0.95\linewidth]{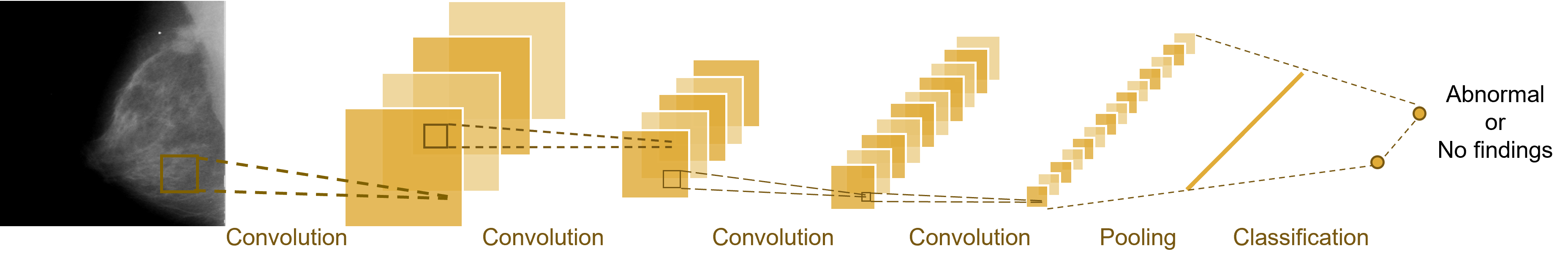}
  \caption{}
  \label{fig:classification}
\end{subfigure}
\begin{subfigure}{0.286\linewidth}
  \centering
  \includegraphics[width=\linewidth]{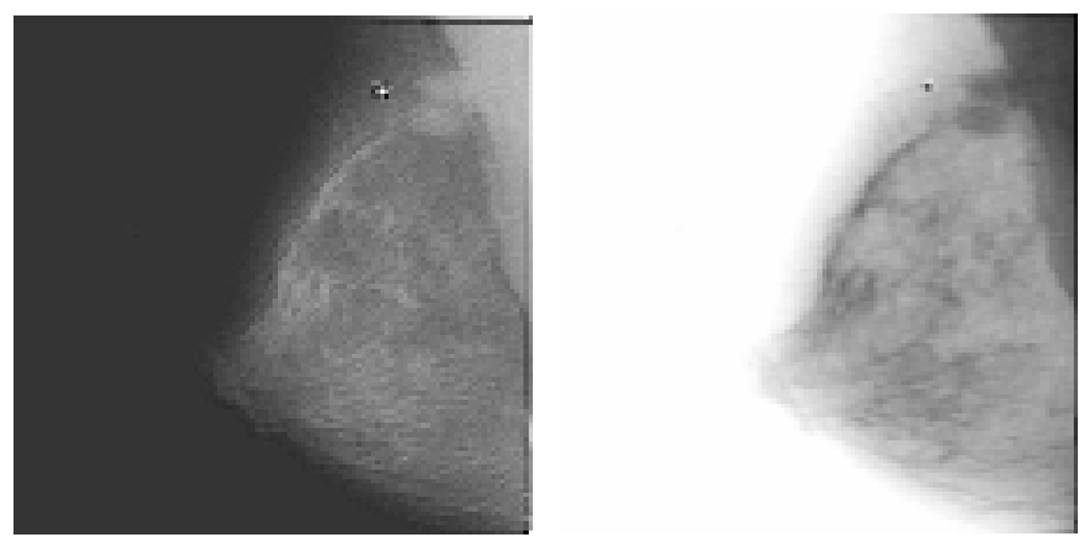}
  \caption{}
  \label{fig:Res18FeatureMap_layer1}
\end{subfigure}
\begin{subfigure}{0.286\linewidth}
  \centering
  \includegraphics[width=\linewidth]{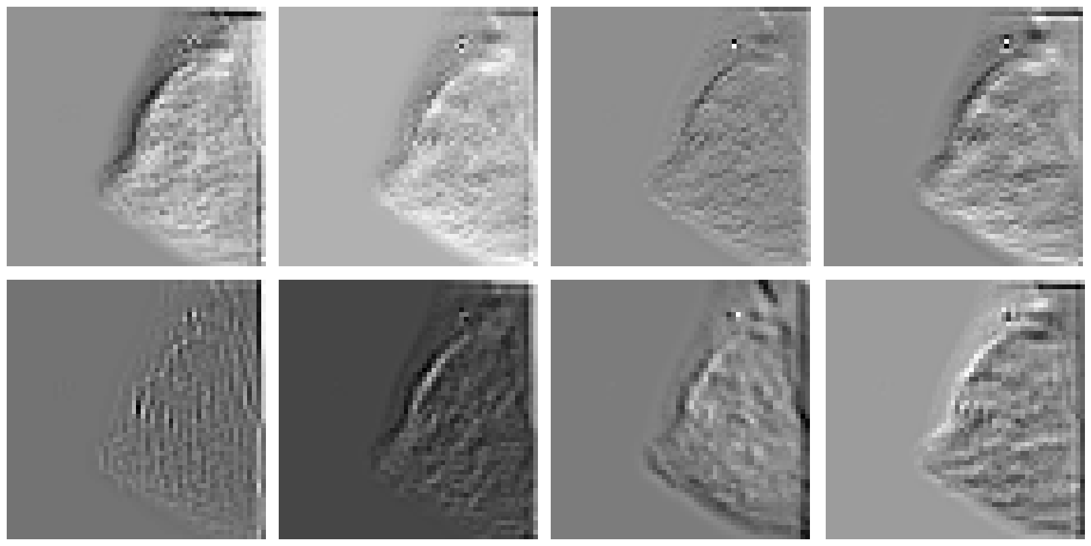}
  \caption{}
  \label{fig:Res18FeatureMap_layer6}
\end{subfigure}
\begin{subfigure}{0.33\linewidth}
  \centering
  \includegraphics[width=\linewidth]{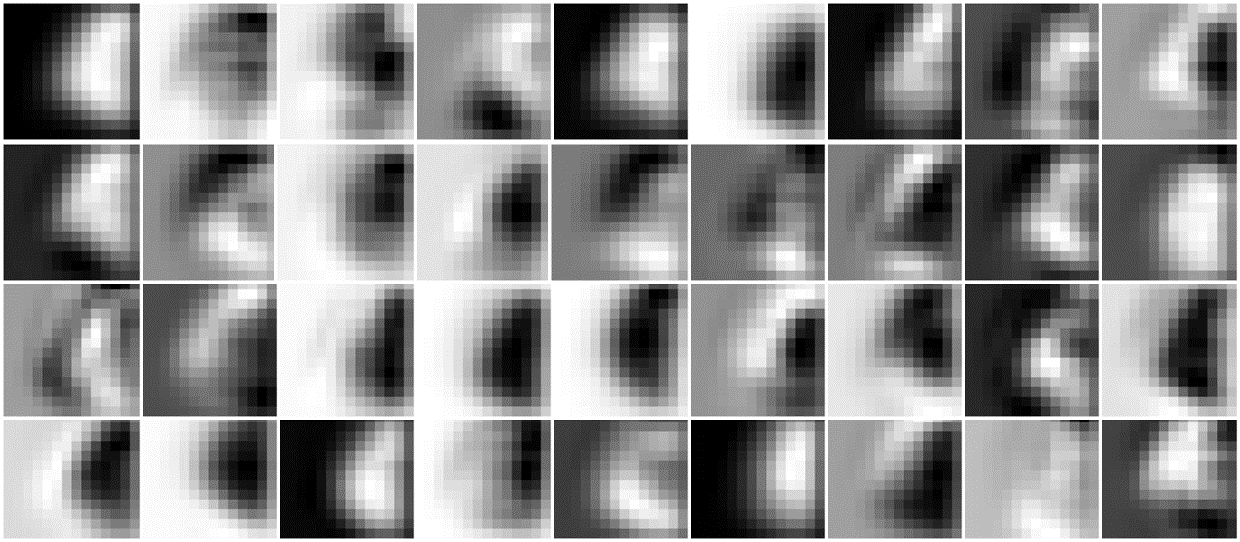}
  \caption{}
  \label{fig:Res18FeatureMap_layer16}
\end{subfigure}
\begin{subfigure}{0.48\linewidth}
  \includegraphics[width=0.9\linewidth]{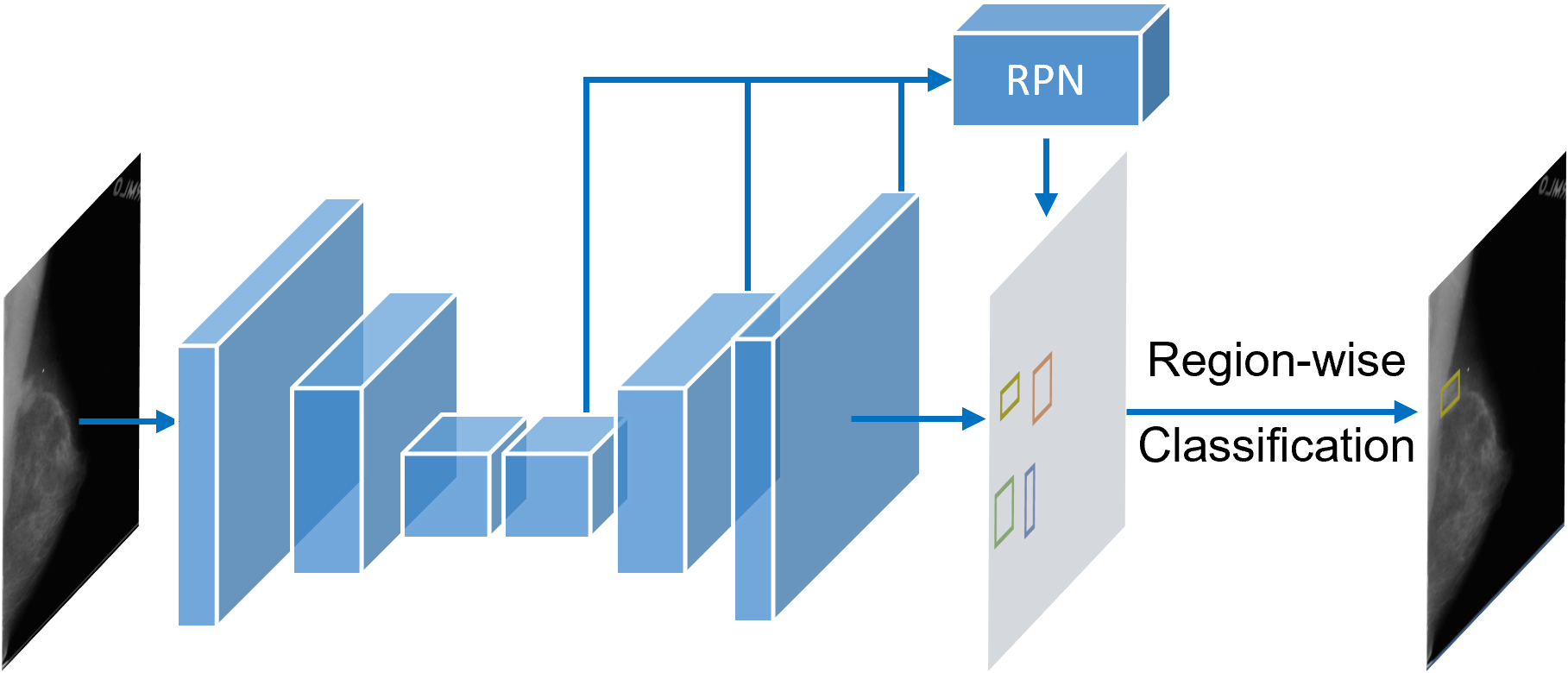}
  \caption{}
  \label{fig:detection}
\end{subfigure} 
\begin{subfigure}{0.48\linewidth}
  \includegraphics[width=0.9\linewidth]{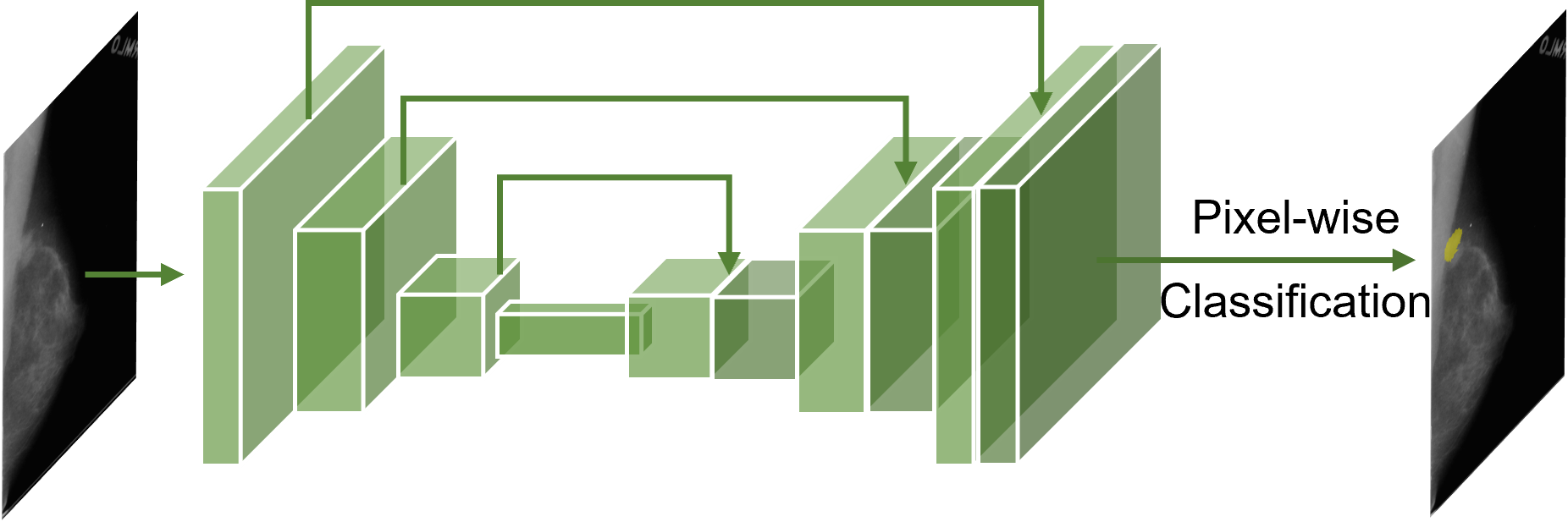}
  \caption{}
  \label{fig:segmentation}
\end{subfigure}
\caption{\textcolor{black}{Brief illustration of deep learning models, taking mammogram as an example. (a) A typical classification network that uses convolutional and pooling to downsample the image while expanding the channels of features. The final feature maps will be pooled into a feature vector, and often a fully-connected layer can be used to conduct the classification based on the feature vector. 
Typical feature maps extracted by a ResNet-18 pre-trained on ImageNet from layers 1, 7, and 17 are shown in (b), (c), and (d), respectively.
(e) A typical detection network. The downsampling workflow often follows the classification network. Then, the feature maps are upsampled, the multi-scale features are fed into a region proposal network (RPN) for region proposal generation, and a region-wise classification is performed to determine the final output. 
(f) A typical segmentation network. The downsampling workflow could follow the classification network. Then, the feature maps are upsampled several times and concatenated with the shallow-layer features. The final results are obtained based on pixel-wise classification on the largest feature map. All the models are optimized with backpropagation \cite{rumelhart1986learning}.}}
\label{Fig:Models}
\end{figure*}

This section will introduce the major deep learning techniques used in breast cancer imaging.
For a more detailed review of deep learning, we refer the readers to \cite{goodfellow2016deep}.
We will first introduce the formulations and some majorly used deep learning models by categorizing breast cancer image analysis into three basic tasks, i.e., classification, detection, and segmentation, according to the output types, \textcolor{black}{and a brief illustration of the deep learning models commonly utilized in each task can be found in Fig. \ref{Fig:Models}.}
We will then introduce the widely applied deep learning paradigms, including supervised learning, semi-supervised learning, weakly-supervised learning, unsupervised learning, transfer learning, and multimodal learning.

\subsection{Classification}
Classification aims to give discrete predictions to categorize the whole inputs, e.g., 1 to indicate that a breast image contains cancer and 0 to indicate that the image does not contain cancer.
A classification model can be regarded as a mapping function $f: X \xrightarrow{} Y$, where $X$ is the domain of images or features and $Y \in \mathbb{R}$ is usually a one-hot representation of the disease existence.
Formally, given $x$ an input, $y$ the target output, and $\hat{y}$ the model output, the classification models are typically optimized by minimizing the cross entropy between $\hat{y}$ and $y$:
\begin{equation}
\label{eq:CE}
    \mathcal{L} = -y{\rm log}\hat{y}
\end{equation}

\textcolor{black}{To model $f$, earlier studies would utilize artificial neural networks (ANNs) \cite{mcculloch1943logical,rosenblatt1958perceptron,rumelhart1986learning} that are constructed by several fully-connected layers and take as input hand-crafted features.}
Convolutional neural network (CNN) \cite{lecun1989backpropagation} gets rid of feature engineering and makes the classification problem on images fully end-to-end.
In 2012, the success of AlexNet \cite{krizhevsky2012imagenet}, a 5-layer CNN powered by graphic processing unit (GPU), kicked off the era of deep learning with its outstripping performance on the ImageNet challenge \cite{deng2009imagenet}.
VGG \cite{simonyan2014very} extended the depth of CNNs with smaller kernels and auxiliary losses.
Residual networks (ResNet) \cite{he2016deep} further deepened CNNs to hundreds of layers and conquered the gradient vanishing problem with skip connections.
Apart from AlexNet, VGG, and ResNet, many other networks like Densely Connected Network (DenseNet) \cite{huang2017densely} and the Inception series \cite{szegedy2015going,szegedy2017inception} have all been widely used in breast cancer imaging.
Recently, vision transformer \cite{dosovitskiy2020image}, a type of deep neural networks that are mostly based on attention mechanism~\cite{vaswani2017attention}, has also shown great potential in image processing.
It is worth mentioning that classification models are often used as a feature extractor for other tasks which will be introduced in the following sections.

\subsection{Detection}
Detection aims to predict region-wise classification results, e.g., drawing a bounding box for a recognized malignancy.
Reusing $f: X \xrightarrow{} Y$ as the mapping function of a detection model, $X$ remains the domain of images, while $Y$ is a set of $\{(b, y)\}$ with $b$ the region and $y$ the corresponding class for that region.
Note that $\{(b, y)\}$ could be an empty set if there are no regions of interest (ROIs) on the image.
The most commonly used formulation of $b$ is a quadruple $\{u, v, w, h\}$, where $u$ and $v$ represent the center of a object box, and $w$ and $h$ represent the weight and height, respectively.
The detection objective is often formalized as sibling tasks containing a region-wise classification loss $\mathcal{L}_{\rm cls}$ and a bounding box regression loss $\mathcal{L}_{\rm loc}$:
\begin{equation}
    \mathcal{L}= \mathcal{L}_{\rm cls} + \lambda\mathcal{L}_{\rm loc},
\end{equation}
where $\mathcal{L}_{\rm cls}$ is commonly formed as a cross entropy loss, $\lambda$ is a loss balancing hyper-parameter, and $\mathcal{L}_{\rm loc}$ is often formalized as a smooth L1 loss as follows:
\begin{equation}
    \mathcal{L}_{\rm loc} = 
    \begin{cases}
        0.5 \times (t - \hat{t})^{2}, &  \textrm{if $|t - \hat{t}|<1$;} \\
        |t - \hat{t}| - 0.5, &  \textrm{otherwise},
    \end{cases}
\end{equation}
where $\hat{t}$ is the model prediction, $t$ is the transformed location representation based on $b$ for regularized regression \cite{girshick2015region}, and $|\cdot|$ represents L1-norm.
There are also other choices for $\mathcal{L}_{\rm cls}$ (e.g., focal loss \cite{lin2017focal}) and $\mathcal{L}_{\rm loc}$ (e.g., the intersection over union loss \cite{rezatofighi2019generalized}).

Object detection models often leverage well-trained classification networks as feature extractors.
To conduct region-wise prediction on the extracted feature maps, a considerable number of deep learning architectures have been proposed, and some of the frequently used models in breast cancer imaging are covered here.
Fast R-CNN \cite{girshick2015fast} extracts and pools the proposals (i.e., the candidate object features) from a pre-trained CNN and conducts the sibling localization and classification tasks.
\textcolor{black}{Faster R-CNN \cite{ren2015faster} introduces the concept of anchors with region proposal network (RPN), which boosts the speed of detection with reference boxes on the feature maps.}
Apart from the mentioned two-stage methods that extract proposals and then conduct classification and regression, one-stage detectors have been proposed to further accelerate the inference speed, with classical representatives such as the YOLO series \cite{redmon2016you,bochkovskiy2020yolov4} and RetinaNet \cite{lin2017focal}.

\subsection{Segmentation}
Segmentation aims to give pixel-wise classification predictions, and the contour of objects can then be delineated.
Taking the segmentation task on 2D images as an example, the input domain is $X\in\mathbb{R}^{W\times H\times C}$ and the output domain is $Y\in\mathbb{R}^{W\times H\times N}$, where $W$, $H$, $C$, $N$ represents the width, height, channel, and the number of classes, respectively.
Typical objectives for medical image segmentation are pixel-wise cross entropy loss and the Dice loss \cite{milletari2016v}:
\begin{equation}
    \mathcal{L} = 1 - \frac{2*\sum_{i=1}^{M}y_{i}\hat{y}_{i} + \epsilon}{\sum_{i=1}^{M}y_{i}^{2}+\sum_{i=1}^{M}\hat{y}_{i}^{2} + \epsilon},
\end{equation}
where $M$ is the total number of pixels, $y_{i}$ is the $i$-th pixel target, $\hat{y}_{i}$ is the $i$-th pixel prediction, and $\epsilon$ is a hyper-parameter for numerical stability.
The target $y$ here is also called the mask of the image.

Segmentation models for medical imaging typically follow an encoder-decoder structure that first downsamples (e.g., via convolution and pooling) the input image into features and then upsamples (e.g., via deconvolution and interpolation) the features to pixel-wise predictions.
FCN \cite{long2015fully} first introduced deconvolution to a VGG classifier for image segmentation.
U-Net \cite{ronneberger2015u} expanded the upsampling process to multiple levels of interpolation and further introduced skip connections between the encoder and decoder paths to enrich details.
Later on, U-Net and its variants \cite{azad2022medical} have almost dominated the field of medical image segmentation.
Further, to separate overlapped instances that belong to the same class, Mask-RCNN \cite{he2017mask} is often adopted, which first detects object proposals on the images and then performs segmentation for each detected object.

\subsection{Deep Learning Paradigms}
\label{Deep Learning Paradigms}

There are diverse options of deep learning paradigms to apply the models to different scenarios, given the availability of the data and labels.

\textbf{Supervised learning} requires all training samples to be labeled exactly in the form of targeted outputs, e.g., masks for the segmentation task or bounding boxes for the detection task.
Supervised learning is the most common form of deep learning, and a large proportion of studies reviewed in this paper fall into this category.
However, deep learning is notoriously data-hungry and labeling medical images is time-consuming, and expertise-depending. Hence, supervised learning may not be the optimal solution for many practical medical image analysis scenarios.

\textbf{Weakly-supervised learning} (WSL) is applied when the given label is not in the format of the targeted output.
For example, using image-level annotations for detection or segmentation.
In breast cancer imaging, the mostly used weakly-supervised learning methods are class activation map (CAM) \cite{zhou2016learning} and multiple instance learning (MIL) \cite{carbonneau2018multiple}.
CAM is often used for rough detection of targeted lesions, which is computed as the feature maps weighted by corresponding gradients.
Higher values on a CAM indicate the regions that contributes more to the final prediction.
MIL treats an input image as a bag of instances (i.e., image patches) which is negative only when all instances are negative.
The goal of MIL is often to develop a bag-level classifier, which is quite a common strategy in processing whole slide pathology images which are of giga-pixel scale.
Like CAM, MIL can also be used to roughly localize the lesions by highlighting the mostly contributed instances.

\textbf{Semi-supervised learning} (SSL) can be regarded as another type of WSL, which enables utilizing a large amount of unlabeled data together with limited labeled data.
Typical SSL methods are based on graph, entropy minimization, pseudo labeling, generative modeling, or consistency learning.
Recently, consistency-based approaches have shown great success in SSL, which inject a regularization on the model that the predictions on different perturbated versions of a model should remain consistent.

\textbf{Unsupervised learning} leverages unlabeled data for model training, often aiming at clustering or dimension reduction.
In the literature of deep learning-based breast cancer image analysis, two major directions in unsupervised learning have gained research attention: generative modeling and self-supervised learning.
The former uses generative methods, such as the generative adversarial network (GAN) \cite{goodfellow2020generative}, to model the data distribution and generate new samples, which is also quite often used in SSL.
Self-supervised learning trains a neural network on the unlabeled images to learn representations for the supervised downstream tasks \cite{jing2020self}.
Self-supervised learning has shown remarkable strength in reducing the requirement of large amount of downstream labeled data, which holds great promise in medical image analysis.

\textbf{Transfer learning} aims to transfer the knowledge learned on a source domain to the target domain, which is especially useful when a target domain does not possess too much data.
A common transfer learning strategy in medical imaging is pre-training the networks on large-scale natural image datasets, such as the ImageNet \cite{krizhevsky2012imagenet}.
Recently, domain adaptation \cite{wang2018deep} and domain generalization \cite{zhou2022domain} have also gained huge attention, which mitigate the gaps between the source domain and the target domain.

\textbf{Multimodal learning} aims to process and summarize information from different views/modalities of a subject.
Taking the examination process of breast cancer as an example, multimodal data from mammogram, ultrasound, MRI to pathology images could be generated and utilized together.
Multimodal data could provide rich and complementary information from macroscope to microscope.
It is worth mentioning that multimodal data could also be yielded with a single type of imaging technique, such as multi-view mammograms or multi-sequence MRIs.
Typically in breast imaging, multimodal learning focuses on the information fusing strategies, which mainly includes early fusion (i.e., fuse data at the input level), joint fusion (i.e., fuse at the feature level), and late fusion (i.e., fuse at the decision level).

\section{Deep Learning Applications in Breast Cancer}
\label{Applications in Breast Cancer}

\textcolor{black}{
We here provide a concise review of deep learning-based applications in breast cancer imaging. 
One may refer to the supplementary where we also provide detailed lists of studies surveyed by us.
}

\subsection{Screening and Diagnosis}
Screening aims to find patients out of the examined cohort, and diagnosis aims to give a more precise description of the patients' status.
Screening is often based on population-scale mammograms, and diagnosis often uses other imaging modalities.
However, there is no clear demarcation between the two tasks in the context of deep learning, as a large number of studies focus on determining malignancies from normal or benign subjects.
We hence introduce the deep learning-based breast cancer screening and diagnosis in a combined section.

\subsubsection{Mammogram-based screening and Diagnosis}

\noindent\textbf{Classification.}
As a routine breast cancer screening approach, mammograms are often studied for binary classification (e.g., malignant vs benign/normal/non-malignant) or three-class classification (e.g., malignant vs benign vs normal).
Studies also tried to distinguish different types of lesions such as mass or architectural distortion, the levels of breast density, or the levels of cancer risks.

\textcolor{black}{The early studies relied on hand-crafted features as inputs to ANNs~\cite{dheeba2014computer,chan2020deep}.}
With their remarkable success in analyzing natural images \cite{krizhevsky2012imagenet}, CNNs were also used as a powerful feature extractor combined with other machine learning classifiers like Support Vector Machine (SVM) and Random Forest~\cite{dhungel2016automated}.

As medical data are often limited in scale, some works \cite{teare2017malignancy,mohamed2018deep} transferred existing networks with pre-trained weights from the ImageNet to the mammogram datasets.
A comparative study on mammogram classification performance of different networks was reported in \cite{adedigba2022performance}.
A comparison of the efficiency of mammogram-based classification (using 2D VGG) and DBT-based classification models (using 3D VGG) was reported by Li et al. \cite{li2020digital}.
Apart from directly using on-the-shelf models, studies further sought more effective transfer learning methods to improve the pre-training learning process and fully utilize the learned knowledge from the pre-training dataset \cite{samala2018breast}.
A comparative study on the pre-training strategies has been reported by Clancy et al. \cite{clancy2020deep}.
To enlarge the training data and learn more robust models, data augmentation and model ensemble have been widely used \cite{sert2017ensemble}.
Several works also attempted to use generative models to enlarge the scale of training data \cite{li2019signed}.

Multiple instance learning (MIL) has played an important role in mammogram-based classification, as the lesions are often sparse \cite{zhu2017deep}.
A line of works focused on developing MIL classifiers with different aggregation strategies to summarize the final feature maps of a CNN model \cite{zhu2017deep,shu2020deep}.
Moreover, Lotter et al. \cite{lotter2021robust} enabled the training of a RetinaNet detector \cite{lin2017focal} with both bounding box annotations and image-level supervision using the MIL strategy.
For 3D DBT classification, some works utilized a 2D classifier to obtain results for each slice and fused the results with MIL for final 3D prediction \cite{yousefi2018mass}.

A large proportion of studies proposed learning patient-level prediction from multiple input images.
On the one hand, multiple views (e.g., the bilateral craniocaudal (CC), mediolateral oblique (MLO) views, etc.) are often generated for more detailed examination.
Different multi-view feature fusion methods have thus been proposed \cite{carneiro2015unregistered,carneiro2017automated}, mostly focusing on developing feature extractors to generate more semantically meaningful representations or fusing strategies to inference final results based on the heterogeneous features.
On the other hand, studies also attempted to combine the information of the previously screened image (i.e., prior mammogram) and the currently examined image (i.e., current mammogram) for malignancy classification \cite{bai2022feature}.
\textcolor{black}{In particular}, Baccouche et al. \cite{baccouche2022early} proposed using GAN to generate a prior image from current mammograms and combining the two images for lesion detection.

Multi-task learning has also been studied to enhance classification performance with extra supervision from other tasks \cite{shams2018deep}.
Notably, multi-task learning and multi-view inputs can both enrich the information provided to the model, and these two strategies can be simultaneously incorporated \cite{wimmer2021multi}.

\noindent\textbf{Detection.}
Lesion detection could specify the location and type of the tumors and further quantify cancer development.

Two-stage cancer detection pipeline was widely used, where hand-crafted \cite{suganthi2012improved} or network-segmented \cite{shen2020multicontext} lesion candidates are fed into a classification network for false positive reduction.
\textcolor{black}{Studies also adopted and modified off-the-shelf end-to-end detectors,} such as Faster-RCNN~\cite{girshick2015fast} and YOLO~\cite{redmon2016you}, which take as input the whole mammography image and output bounding box coordinates for lesions with scores indicating the likelihoods of different lesion types \cite{buda2021data}.
Li et al. \cite{li2022architectural} also proposed a cancer detection method for 3D DBT by aggregating the 2D predictions into 3D results.

Multi-view methods are also frequently studied for breast lesion detection.
Based on the same feature extractor backbone for each view, Liu et al. \cite{liu2021compare} compared and fused the features from different views to improve the detection performance on top of Mask-RCNN \cite{he2017mask}.
Graph-based reasoning was also integrated into the multi-view detection framework with graph nodes generated by pseudo 
landmarks \cite{liu2021act}.
Recently, transformer-based detector was also used for multi-view mass detection \cite{zhao2022check}.
On the other hand, Yang et al. \cite{yang2021momminet} proposed to fuse the features of different networks, where each network is designed for a specific view.

It is also noteworthy that the feature maps extracted from a classification network can also be used to localize the lesions by CAM or the attention mechanism in a weakly supervised manner.
However, quantitative evaluations are often lacking for these studies, and the attention maps are often used for qualitative interpretation purpose \cite{zhu2017deep,shu2020deep}.
A recent study also proposed utilizing intermediate features and CAM as pseudo labels to train a detection model \cite{liang2020weakly}.

\noindent\textbf{Segmentation.}
Segmentation provides contour delineation for a more detailed description of the lesion, which often requires a considerable amount of pixel-wise annotations by experienced radiologists for model training. 

There are a few studies on mass segmentation from mammograms, based on modified structures or losses with FCN \cite{zhu2018adversarial}, conditional GAN \cite{mirza2014conditional,li2021sap}, or U-Net-alike structures \cite{gudhe2022area}.
A comparative study on different segmentation models was conducted in \cite{ribeiro2022deep}.

\noindent\textbf{Multiple Tasks.}
Some studies tried to integrate modules for different tasks to establish multi-functional computer-aided diagnosis (CAD) frameworks.

A straightforward way to a multi-task CAD system is training a set of task-specific models \cite{yi2021deepcat}, where the studies are often based on existing solutions for each task.
Object detection networks, such as Faster R-CNN, Mask R-CNN and YOLO, have also been adopted to conduct both classification and detection due to their multi-task learning nature \cite{min2020fully,baccouche2022early}.
As mentioned, classification models could be used to extract detection results in a weakly-supervised manner as well \cite{hwang2016self,bakalo2019classification}.
Moreover, unified multi-task models have been proposed in the literature, e.g., combining classification with detection \cite{sainz2020multi} or segmentation \cite{li2021dual}.

\noindent\textbf{Other Tasks.}
Many studies aim for more than developing specific methods for the aforementioned basic tasks.

A comparative study of mammogram classification performance between a deep learning model and 101 radiologists has been reported by \cite{rodriguez2019stand}, where the model showed comparable performance to the average of radiologists.
Later on, a handful of studies have shown that deep learning models could help improve the radiologists' performance \cite{wu2019deep,rodriguez2019detection,pinto2021impact}.
Recent study also reported that deep learning surpassed the performance of traditional risk prediction algorithms \cite{lehman2022deep}.
Moreover, deep models have been demonstrated capable of screening mammograms based on large or even international populations \cite{laang2021identifying,mckinney2020international} and finding out high-risk subjects for further examination by radiologists \cite{rodriguez2019can}.

Despite achieving expert-comparable or expert-surpassed accuracy, recent studies reported significant performance drop when applying deep learning models to external testing mammograms \cite{wang2020inconsistent,hsu2022external}.
An increasing number of works attempted to improve the robustness of deep learning models for mammograms in the aspects of noisy label \cite{dgani2018training}, adversarial attacks \cite{zhou2021machine}, external domain generalization \cite{li2021domain}, and privacy preserving under federated learning \cite{roth2020federated}.
Efforts have also been devoted to make the models more interpretable \cite{wang2022knowledge}.

To reduce potential side effects caused by extra imaging process, recent studies also made efforts on mammograms synthesis  \cite{jiang2021TMIsynthesis}.

\subsubsection{Ultrasound-based Diagnosis}

\noindent\textbf{Classification.}
Breast ultrasound (US) images are often used for malignant lesion identification, lymph node metastasis estimation, or breast risk prediction, as an appealing non-invasive alternative to traditional invasive approaches.
Similar to the studies on mammograms, well-validated classification networks like VGG, ResNet, DenseNet, etc., have also been widely adopted in breast ultrasound classification.
Some studies cropped suspicious lesions in US images as the regions of interest (ROIs), which were later fed into the CNNs for malignancy classification~\cite{cheng2016computer,moon2020computer} and metastasis estimation \cite{zheng2020deep,zhou2020lymph,guo2020deep}.
However, a prominent drawback of ROI-based analysis is that lesion regions should be manually cropped in advance, which not only increases the annotation burden of the experts but also impedes the flexibility of these methods in real-world applications. 
Instead of using manual crops, Lee et al. \cite{lee2021axillary} first leveraged a Mask R-CNN model to detect and segment lesions and then used a DenseNet121 model for auxiliary lymph node metastasis status prediction based on the extracted peritumoral tissues.
On the other hand, researchers also attempted to analyze the whole US image without lesion candidate detection in the first place~\cite{xing2020using,huang2021aw3m,lu2022safnet}.
It is worth noting that with transfer learning and data augmentation techniques, the whole US image-based works can achieve comparable performance to the studies based on carefully extracted ROIs.

In clinical routine practice, radiologists normally make diagnostic decisions based on a comprehensive evaluation of US images in different views and \textcolor{black}{a combination of} different modalities such as B-mode, color Doppler, and elastography images.
Therefore, studies have also attempted to fuse the complementary information from multi-view or multimodal US images, where feature-level fusion \cite{qian2021prospective,huang2021aw3m} and output-level (predicted probability for each view) fusion \cite{tanaka2019computer} have both been explored.
 
To alleviate the overfitting problem resulting from data deficiency in training, researchers applied various strategies, such as using ImageNet pre-trained models, data augmentation \cite{han2017deep}, or transfer learning from other related tasks \cite{hijab2019breast}. 
Moreover, GANs have also been used for generating synthetic data \textcolor{black}{as a surrogate of data augmentation}, which demonstrated excellent performance in the task of classification~\cite{pang2021semi}. 

To efficiently leverage limited annotated data and a large amount of unlabeled data for training, unsupervised learning~\cite{zhang2020birads} and semi-supervised learning techniques~\cite{wang2021deep} were also explored to enhance the discrimination capability of the model, based on techniques like autoencoder, Mean Teacher~\cite{tarvainen2017mean}, and  Virtual Adversarial Training (VAT)~\cite{miyato2018virtual}.

\noindent\textbf{Detection.}
Detection of lesions is also clinically preferable for breast US diagnosis. 
Many studies used CNNs for candidate classification and false positive reduction after a lesion candidate extraction process \cite{yap2017automated,chiang2018tumor,wang2019deeply}.
A comparative study on a variety of state-of-the-art detection networks such as Faster R-CNN and YOLO, and classification networks such as AlexNet and DenseNet were systematically reported in~\cite{cao2019experimental}. 
To further incorporate the large amount of data annotated at the image level, Shin et al. \cite{shin2018joint} proposed a joint weakly- and semi-supervised network based on the multi-instance learning scheme.

\noindent\textbf{Segmentation.}
Segmentation of important ROIs from US images, such as the tumor region, major functional tissues, and breast anatomic layers, aims to provide more fine-grained and quantitative information to clinicians. 
Automated object segmentation from ultrasound images is quite challenging owing to speckle artifacts, low contrast, shadows, blurry boundaries, and the variance in object shapes.
Recently, deep learning-based approaches, particularly FCN and U-Net, have been successfully applied to this field~\cite{xing2020lesion,gomez2020comparative}. 

To enhance the confidence of the hardly-predicted boundary, a number of strategies were designed in previous works~\cite{ning2021smu,xue2021global,huang2022boundary}. 
For instance, Xue et al. \cite{xue2021global} developed a breast lesion boundary detection module in shallow CNN layers to embed additional boundary maps of breast lesions for obtaining the segmentation result with high-quality boundaries. 
Other boundary-aware modules were also proposed to achieve more precise segmentation in the confusing and ambiguous boundary areas~\cite{ning2021smu,huang2022boundary}. 

Different variants of attention mechanisms were used in conjunction with the deep learning model for US segmentation, such as paying spatial attention to most specific regions in the US images, and weighting the feature channels that have different semantic information~\cite{ferreira2022comparative}.
In addition, saliency maps that highlight visually salient regions or objects were also explored to strengthen the network’s attention to the region of interest in US mages and help to boost the segmentation performance of the model~\cite{ning2021smu}.

Similar to the classification studies, GANs were widely applied to synthesize training data to augment the training set for improving the segmentation results~\cite{xing2020lesion,zhai2022ass}.

\noindent\textbf{Multiple Tasks.}
A complete CAD system often requires multiple functions, such as lesion detection and classification~\cite{shin2018joint,cao2019experimental,tanaka2019computer} or lesion segmentation and classification~\cite{liao2019automatic},
either in a manner of sequential modeling or multi-task learning. 
\textcolor{black}{Some accomplished such goal in two steps}, i.e., first detecting or segmenting the lesion regions, and then classifying the detected or segmented areas into benign or malignant classes~\cite{shin2018joint,tanaka2019computer,lee2021axillary}. 
\textcolor{black}{By contrast, there} were also methods proposed to conduct multiple tasks simultaneously in an integrated framework~\cite{zhou2021multi}. 
For instance, Zhou et al. \cite{zhou2021multi} proposed a multi-task learning framework for joint segmentation and classification of tumors in 3D automated breast ultrasound images. It is composed of two sub-networks: an encoder-decoder network for segmentation and a light-weight multi-scale network for classification. 

Weakly-supervised segmentation has also been explored in breast US~\cite{kim2021weakly}, where CAM-based methods were often applied in a post-hoc manner to highlight the most discriminative regions discovered by the model. 

\noindent\textbf{Other Tasks.}
Some studies explored novel modalities other than conventional US images, such as the contrast-enhanced ultrasound videos~\cite{chen2021domain} and automated whole breast ultrasound images~\cite{lei2018segmentation,chiang2018tumor,wang2019deeply,zhou2021multi,zhou2021cross}, for lesion classification, detection, and segmentation, where 3D CNNs were often built to leverage the rich \textcolor{black}{temporal} or spatial information for more robust learning~\cite{chiang2018tumor,chen2021domain}.

\textcolor{black}{Observer study} by radiologists is quite important to validate the potential of DL methods in real-world applications.
There are also many works demonstrating that the deep learning approaches have achieved expert-level performance, showing promises in commercial applications~\cite {qian2021prospective}.

\textcolor{black}{To improve model robustness against noise}, Cao et al. \cite{cao2020breast} built a noise filter network to prevent classification models from overfitting the noisy labels. 
Zou et al. \cite{zou2021robust} developed an end-to-end noisy annotation tolerance network for robust US image segmentation.

\subsubsection{MRI-based Diagnosis}

As the most sensitive radiological modality for breast cancer detection, MRI examinations are applied to fine-grained diagnosis of breast cancer and provide more detailed preoperative guidance for treatment planning. 
We here generally categorize previous studies into classification, segmentation, detection, and other tasks.

\noindent\textbf{Classification.}
Classification of breast MRI could be \textcolor{black}{categorized} from three perspectives according to their purpose: First, the basic screening task by detecting the absence of lesions which helps physicians to dismiss the normal MRI examinations.
Second, a binary classification distinguishing malignant tumors from benign ones for follow-up treatment regimens occupies a large part of research.
Third, fine-grained classification like predicting molecular subtypes, BI-RADS, and metastasis also facilitate detailed and further diagnosis.
Currently, a large proportion of studies are based on DCE-MRI that takes several MR images at different time points after injection of contrast agent.
Ultrafast MRI, which takes a short scan duration within several seconds, has also gained large research interest as patients with motion sensitivity and emergency situations could get accurate radiology examinations from ultrafast MRI promptly and get timely treatments.
Moreover, learning from multi-parametric MRI enables the fusion of more comprehensive information from multiple modalities, which has also triggered exploration and improvements.

Different deep learning architectures have been proposed regarding the diverse MRI sequences.
Compared with other sequences, DCE-MRI further possesses temporal information provided by the contrast agents. 
Long short-term memory (LSTM)-based \cite{feng2020knowledge} or convolutional LSTM-based \cite{zhang2021prediction} networks were applied to exploit the temporal features involved in DCE MRI sequences. 
\textcolor{black}{To lower the computation cost of 3D MRI}, maximum intensity projection (MIP), which projects the voxels throughout the volume onto a 2D image, has also been applied in the DCE series \cite{antropova2018use,fujioka2021deep,hu2021improved}. 
For instance, from a comparison study based on MIP and central slides of different sequences, Antropova et al. \cite{antropova2018use} witnessed the effectiveness and advantages of time and space-saving of MIP techniques.

Learning from the multi-parametric MRI combining multiple sequences and integrating their corresponding advantages is another research focus. 
For instance, Hu et al. \cite{hu2020deep} implemented three different fusion strategies: image fusion of DCE MIP and \textcolor{black}{center slice} of T2-weighted modality, feature fusion of CNN-based deep features, or classifiers output fusion for two training branches of modalities. 
It's worth noting that feature fusion was found performed significantly better in classifying between benign and malignant lesions.
Ren et al. \cite{ren2022convolutional} compared five models based on different MRI modalities on the task of axillary lymph node metastasis prediction, where the combination of DCE + T2 inputs was found to perform best.
Some studies also combined deep features with hand-crafted radiomics features \cite{whitney2019comparison} to improve the classification performance.

Apart from network architecture design, other techniques have also been embedded in MRI-based CAD systems.
Inspired by transfer learning, using pre-trained networks \cite{jing2022using,hu2021improved} could solve the data shortage to some extent and speed up the convergence. 
Also, ensemble learning frameworks \cite{sun2021prediction,rasti2017breast} were introduced to reduce the model uncertainty. 
Rasti et al. \cite{rasti2017breast} designed multiple gating networks sharing the same input and fused the outputs at last.
Sun et al. \cite{sun2021prediction} predicted the molecular subtypes (luminal and non-luminal) based on the ensemble outputs from three sub-models trained with different post-contrast sequences. 

Weakly-supervised and unsupervised learning have been studied to tackle the scarcity of labels.
Liu et al. \cite{liu2022weakly} classified the benign and malignant tumors from the whole slides instead of the targeted region of interest with ResNet-based networks. 
Sun et al. \cite{sun2022transfer} utilized transfer learning strategies from the source domain with unsupervised pre-training on DCE-MRI containing benign and malignant cohorts for predicting molecular subtypes.

\noindent\textbf{Detection.}
Compared with other modalities, studies on breast MRI-based detection is of a relatively smaller scale.
Dalmics et al. \cite{dalmics2018fully} took the lesion candidate patch and its contralateral patch as inputs for lesion classification.
Maicas et al. \cite{maicas2017deep} incorporated reinforcement learning-based Deep Q network \cite{mnih2015human} with an attention mechanism for breast lesion detection, which showed accurate localization while saving more inference time. 
Ayatollahi et al. \cite{ayatollahi2021automatic} modified a 3D RetinaNet for small breast lesion detection on ultrafast DCE-MRI.

\noindent\textbf{Segmentation.}
U-Net-based architectures \cite{yue2022deep,vidal2022u,galli2021pipelined} are the most commonly used structures for breast tumor segmentation.
In addition, considering the temporal information and physiological inheritance involved in the DCE-MRI, Three Time Points (3TP) approach was introduced to help quantize the intensity change of the breasts before, during, and after injection of the contrast agent.
For example, Vidal et al. \cite{vidal2022u} fused the outputs from independent U-Net branches segmenting different series combinations such as 3TP series and full series. 
Galli et al. \cite{galli2021pipelined} extracted the 3TP slices after breast masking and motion correction for lesion segmentation. 
Moreover, some work \cite{zhang2018hierarchical} conducted multi-stage coarse-to-fine segmentation. 

\noindent\textbf{Multiple Tasks.}
For a more complete CAD system, lots of works implement the combination of tasks simultaneously. 
Studies \cite{zhu2022development,parekh2020multiparametric} implemented the segmentation and classification tasks in one pipeline with sequential order for more diagnosis analysis on the segmented lesions. 
Zhu et al. \cite{zhu2022development} utilized VNet and Attention U-Net to segment lesions from DCE and DWI, and ResNet was then used for benign and malignant classification based on the segmentation outputs. 
Parekh et al. \cite{parekh2020multiparametric} utilized stacked sparse autoencoder networks for segmentation of intrinsic tissue signatures, which was followed by an SVM classifier for classification. 

Some studies combined classification and weakly-supervised detection \cite{zhou2019weakly,luo2019deep} to achieve a comprehensive diagnosis. 
Zhou et al. \cite{zhou2019weakly} implemented the classification task on 3D densely connected networks and localized the lesions with Class Activation Map and conditional Random Dense Conditional Random Field. 
Luo et al. \cite{luo2019deep} proposed Cosine Margin Sigmoid Loss for learning cancer malignancy classification and Correlation Attention On-the-shelf models, such as Faster R-CNN, were also used for ROI localization first and combined with custom CNNs for lesion classification \cite{chen2022deep}. 

\noindent\textbf{Other Tasks.}
Breast density estimation by 3D CNN-based regression \cite{van2020volumetric} showed benefit in helping breast cancer risk prediction. 
Predicting the biomarkers, such as Ki-67 status, can help indicate the development of breast cancer. 
For example, Liu et al. \cite{liu2021preoperative} fused the deep features based on transfer learning CNN frameworks from multi-parametric MRI and \textcolor{black}{predicted} the Ki-67 status with a multilayer perceptron classifier. 
Moreover, to increase \textcolor{black}{the accessibility} of breast MRI, Chung et al. \cite{chung2022deep} generated simulated multi-parametric MRI based on 3D fully convolutional networks and validated its quality by comparing it with real scans.

\subsubsection{Digital Pathology Images-based Diagnosis}
\sloppy
Pathology image-based diagnosis plays an irreplaceable role as the ``gold standard" \textcolor{black}{in cancer characterization.}
Deep learning-based breast pathology diagnosis has \text{black}{flourished} in the past decade \textcolor{black}{along with an increasing number of} publicly available datasets.
Here, we categorize the studies into classification, segmentation, detection, and other tasks.

\noindent\textbf{Classification.}
The cancer type/grade assignment is an essential task in breast pathology image analysis.
Conventional methods for breast pathology classification are based on hand-crafted features qualitatively designed by domain experts, such as the spatial distribution, arrangement, and individual types of discrete tissue elements or primitive shapes (e.g., nuclei, lymphocytes, or glandular structures).
In the past few years, deep learning methods have been widely used to \textcolor{black}{automate this process for a higher accuracy}~\cite{hatipoglu2014classification}.
In parallel with breakthroughs in deep learning algorithms, the availability of large-scale datasets has also been a critical factor for the success of deep learning methods.
One milestone is the CAMELYON16 dataset~\cite{bejnordi2017diagnostic}, which contains 400 histopathology whole slide images (WSIs) of lymph node sections.
Wang et al. \cite{wang2016deep}, the winner of the CAMELYON16 challenge, adopted various DL models for the patch-wise classification task.
The patch-wise classification results were further aggregated to obtain the geometrical and morphological features of the whole slide image (WSI).
Then the embedded features were used to classify the WSIs into metastasis or negative findings via a Random Forest classifier.
\textcolor{black}{Following the success, many studies have been conducted to improve the performance of patch-wise classification}, including the modification of model architectures~\cite{bejnordi2017context,kong2017cancer}, the use of different data preprocessing techniques~\cite{cruz2017accurate}, and attention mechanism~\cite{ilse2018attention}.
\textcolor{black}{A more recent challenge summarized multiple solutions for quantitative tumor cellularity assessment in breast cancer histology images following neoadjuvant treatment \cite{petrick2021spie}.}

\textcolor{black}{However, a major drawback of these studies is that the patch-wise classification methods usually require extensive manual annotation at the pixel/patch level by expert pathologists, which is time-consuming and labor-intensive for WSIs that are of gigapixel scales.} 
For example, the CAMELYON16 dataset consists of 270 WSIs at 40$\times$ magnification, with roughly the same number of pixels as the \textcolor{black}{entire} ImageNet dataset~\cite{deng2009imagenet}, which was recognized as one of the largest datasets in the field of computer vision. \textcolor{black}{How to leverage coarsely annotated data (e.g., globally labeled WSIs) to effectively train a well-performed model is much more preferable as WSI-wise analysis is often the ultimate goal of pathologists.}

Thus, recent studies have been conducted to develop methods that only require the WSI-level annotations, which are much more convenient and cost-effective for real-world clinical pathological practice.
For example, Campanella et al. \cite{campanella2019clinical} proposed a deep learning method that only requires WSI-level annotations, which formulates the WSI classification task as an example of the multiple instance learning (MIL) problem.
Typically, the MIL-based deep learning method consists of two stages:
in the first stage, a deep neural network is used to extract the features of the instances;
in the second stage, the instance-level features are aggregated to obtain the bag-level features \textcolor{black}{that are fed into a bag-level classifier to yield the final prediction for bags.}
The follow-up studies tried to improve the MIL-based WSI classification method by various strategies, including enhancing the instance representation~\cite{valieris2020deep,saini2022vggin}, extracting more discriminative features~\cite{campanella2019clinical,mehta2022end,wang2022label}, and improving the aggregation strategy~\cite{li2021dualwsi,ilse2018attention,lu2022slidegraph}.
\textcolor{black}{How to improve the information connection between the feature extractor and the aggregator to obtain more discriminative whole-slide level representation remains an open problem, not only for breast pathology image analysis but also for all other DL studies based on WSIs. 
Recent studies have made efforts with techniques such as coupled iterative training \cite{wang2023iteratively} and end-to-end training with huge hardware (GPU) support \cite{wang2023image}.}

\noindent\textbf{Detection.}
Mitosis detection is a representative research field in pathology image analysis.
The density of mitosis is used to assess the cell proliferation activity, which is a key factor for the prognosis of breast cancer~\cite{balkenhol2019deep}.
Recent DL methods on automatic mitosis detection can be divided into three categories: 1) object detection-based methods; 2) two-stage methods; and 3) pixel-wise segmentation methods.
The object detection-based methods~\cite{sebai2020maskmitosis} mainly employed popular detection frameworks such as Mask R-CNN.
The localization and classification of mitosis were performed simultaneously.
The second line of methods consists of two stages.
In the first stage, the mitosis candidates were detected by the object detection methods, including Mask R-CNN~\cite{li2018deepmitosis}, Mask R-CNN~\cite{sohail2021mitotic}, RetinaNet~\cite{almahfouz2021domain}, etc.
In the second stage, the mitosis candidates were classified into mitosis or non-mitosis~\cite{li2018deepmitosis,sohail2021mitotic,lei2020attention}.
The follow-up methods further improved the detection performance in various aspects, including more representative features~\cite{li2018deepmitosis}, network architectures~\cite{sohail2021mitotic,lei2020attention}, and training strategies~\cite{tellez2018whole}, etc.
\textcolor{black}{The third type of method achieve the mitosis detection task via fine-grained segmentation}, which will be introduced in the following section.

\noindent\textbf{Segmentation.}
According to the Nottingham Grading System~\cite{elston1991pathological,rakha2010breast}, the grading of breast cancer is based on the assessment of three morphological features: 1) degree of tubule or gland formation, 2) mitotic count, and 3) nuclear pleomorphism.
Thus, segmentation of glands and nuclei is a fundamental yet crucial task in breast pathology image analysis.
For gland segmentation, existing DL-based methods typically adopt U-Net and its variations~\cite{su2015region}. One of the most popular research foci is to explore the boundary information to boost the segmentation performance~\cite{chan2019histosegnet}.

In parallel with the progression of gland segmentation in breast pathology, segmentation of nuclei has also been well studied, which is used to extract the morphological features of the nuclei, such as size, shape, and texture.
The morphological features of the nuclei assess the nuclear pleomorphism, which can be used to predict the diagnosis and prognosis of breast cancer~\cite{lu2018nuclear}.
In DL-based nuclei segmentation methods, the main challenge is to obtain accurate segmentation results for nuclei with complex shapes and overlapping.
To address this challenge, studies proposed to use multi-task learning~\cite{zhao2021net,graham2022one}, multi-scale learning~\cite{wan2020robust}, and adversarial learning~\cite{mahmood2019deep}.
Further, the other methods exploited information of the nuclear contour within the training stage. 
The most straightforward way is to simultaneously predict the contour and the segmentation mask~\cite{kumar2017dataset}.
\textcolor{black}{In this manner, the} instance segmentation results can be obtained by the post-processing of the contour and the segmentation mask.
The follow-up methods boosted the performance in the aspects of pre-training~\cite{xie2020instance}, data augmentation~\cite{lin2022insmix}, network architectures~\cite{zhou2019cia}, and loss functions~\cite{naylor2018segmentation,graham2019hover}.

Another challenge of nuclei segmentation is the scarce manual annotation.
One WSI can contain tens of thousands of nuclei, which makes the manual annotation of the nuclei segmentation \textcolor{black}{infeasible}.
To address this challenge, some studies aimed at developing methods that only require weak annotations, such as scribbles~\cite{lee2020scribble2label} or even point annotations~\cite{qu2020weakly,lin2023nuclei}.
Based on the size and shape assumptions of the nuclei, existing weakly-supervised methods typically encode the morphological priors into the weak annotations, transforming the weak annotations into the coarse pixel-wise annotations, such as pseudo edge maps~\cite{yoo2019pseudoedgenet}.
Further, the following studies proposed various techniques to eliminate the bias of the inaccurate and incomplete coarse annotations, such as self-training~\cite{qu2020weakly}, co-training~\cite{lin2023nuclei}, and multi-task learning~\cite{yoo2019pseudoedgenet}.

\noindent\textbf{Other Tasks.}
Due to the significant variance of the staining and the imaging conditions (e.g., slide preparation and microscope scanning), the DL-based breast pathology image analysis methods could suffer from the domain shift problem.
Domain shift refers to the data heterogeneity between the source and the target domains~\cite{challen2019artificial}.
Existing methods for mitigating the domain shift problem of pathology images can be divided into three categories: 1) data augmentation, 2) domain adaptation, and 3) domain generalization.
Data augmentation is a common technique to enhance the robustness of the DL models by increasing the diversity of the training data.
In breast pathology analysis, the commonly used data augmentation methods is color distortion on both the RGB channels~\cite{faryna2021tailoring} and the HSI (hematoxylin, eosin, and residual) channels~\cite{tellez2018whole}. 
For domain adaptation, the most representative method is adversarial learning.
For example, some \cite{mahmood2019deep,de2021deep} proposed to use the generative adversarial networks to map the images from the source domain to the target domain.
The domain generalization methods aim at learning domain-invariant features, which can be used to generalize the models to unseen domains.
The typical methods for domain generalization in breast pathology are feature alignment~\cite{alirezazadeh2018representation} and domain-invariant feature learning~\cite{lafarge2017domain}.

Recently, as one type of domain adaption, virtual staining techniques attract largely attention in breast pathology analysis.
For instance, immunohistochemical (IHC) staining reflects protein expression, which is vital for diagnosing cancers, histological classification, grading, staging, and prognosis of tumors.
However, the IHC staining procedure is costly, laborious, and time-consuming.
To complete the diagnosis, virtual HER2 IHC staining methods~\cite{rivenson2018deep} were proposed to transform autofluorescence microscopic images of breast tissue sections into bright-field equivalent microscopic images, matching the HER2 IHC staining that is chemically performed on the same tissue sections.

In addition, \textcolor{black}{observer studies} by pathologists are also a crucial research issue in clinical practice.
Several studies have shown the significant effectiveness of deep learning models compared to that of doctors and pathologists in various applications, such as lymph node metastases detection \cite{bejnordi2017diagnostic} and pathology-based diagnosis~\cite{noorbakhsh2020deep}.

\subsection{Treatment Response}

Different regimens targeting at breast cancer have been well proposed for patients with pertinence based on the specific subtype of tumor, anatomic cancer stage, personal preferences and toxicity risk etc. \cite{waks2019breast}. 
Assessing treatment response is of significance for monitoring the progression of cancer and therapeutic effects, which could help implement further clinical decisions and improve patients' outcomes with personalized treatment plans.

Most of the studies focus on Neoadjuvant Chemotherapy (NAC) \cite{charfare2005neoadjuvant} response prediction.
Neoadjuvant treatment, or preoperative treatment, has become a safe and often effective therapeutic choice for larger primary and locally advanced breast cancer \cite{thompson2012neoadjuvant}, and NAC is one of the most mainstream chemotherapies presently \cite{charfare2005neoadjuvant}. 
Apart from the imaging modalities we introduced before, Computed Tomography (CT) \cite{choi2020early,qi2022multi} is also studied for treatment response prediction.

\noindent\textbf{Classification.}
A qualitative metric to assess NAC is whether the patients achieve pathological complete response (pCR) or not \cite{von2012definition}, which is demonstrated as an indication of high disease free survival rate based on the absence of cancer cells combined with involvement of lymph nodes after treatment course \cite{cortazar2014pathological,liang2022machine}. 
Therefore, a binary classification problem is formulated to identify the pCR and non-pCR of NAC treatment, which could help physicians determine further therapeutic plans. 

Directly applying CNNs models \cite{choi2020early,liu2020novel,ha2019prior,farahmand2022deep} for classification of NAC response is one of the most common and straightforward strategies. 
Moreover, replacing the last fully connected layers of CNN models with other robust conventional classifiers such as Random Forest or Support Vector Machine \cite{comes2021early,li2022deep,jiang2021ultrasound} also attracts a lot of attention. 
The CNN modules are considered as feature extractors providing image representations, which are often combined with other hand-crafted features or clinical information for training conventional classifiers at last. 
Non-imaging data sources such as pathological records as supplementary to imaging representations could provide more comprehensive information for better performance. Modified and improved network architectures could also achieve NAC response performance. Taleghamar et al. \cite{taleghamar2022deep} concatenated features from two branches consisting of modified ResNet and modified residual attention network and output classification with a fully connected network. Qi et al. \cite{qi2022multi} proposed a modified 3D MultiResUnet with Gradient-weighted class Activation Map (Grad-CAM) which could mark the interesting regions during the training process, and show its improvements compared with conventional radiomics analysis. 

The multimodal data has also sparked an in-depth exploration of the research on combining with multimodal learning. 
Joo et al. \cite{joo2021multimodal}  took multi-parametric MRI as inputs such as T1 weighted, T2 weighted and clinical information in parallel and concatenated features at last. 
In addition, clinical information, molecular information, kinetic information, etc., could also be fused into DL frameworks based on multi-stage fusion strategies \cite{li2022deep,aghaei2015computer,peng2022pretreatment,jiang2021ultrasound}. 
Notably, some studies \cite{li2022deep,peng2022pretreatment} fused the molecular types information which has been known as having correlation with NAC response results with imaging features based on CNN models. 
Also, the handcrafted features extracted based on pathological knowledge with conventional algorithms such as histograms \cite{qi2022multi,jiang2021ultrasound} were also incorporated and fused into the image based deep learning frameworks.

Unlike the diagnostic tasks which mostly take one-phase examinations, a large part of source data for predicting treatment utilizes multiple-phase scans across cycles of treatments. Researchers attempt to highlight such temporal information and differences before and after treatment. Methods of independently inputting images of different stages gained many explorations \cite{xie2022dual,tong2022dual}. For example, Xie et al. \cite{xie2022dual} proposed a framework of dual-branch CNN-based models with inputs of images extracting before and after NAC treatments, and then implemented feature fusions from convolutional blocks. 
Similarly, Tong et al. \cite{tong2022dual} proposed dual-branch transformers for NAC response prediction on US images. 
Siamese architectures-based CNNs \cite{liu2022early,byra2020early} were also explore to capture the differences between images before and after treatment cycles. 
In addition, Recurrent Neural Networks (RNN) were used for capturing the information across temporal dimensions and achieved better prediction outcomes on long-term sequential treatment cycles \cite{byra2022prediction}. For multiple cycles of NAC treatments, Gu et al. \cite{gu2022deep} constructed a deep learning pipeline for step-wise prediction of different stages of treatments. 
A comparison study of breast DCE-MRI contrast time points for predicting NAC response has been reported by Huynh et al. \cite{huynh2017comparison}.

Transfer learning decreases the need for a large amount of new data and speeds up the training process on similar tasks. 
Some \cite{li2022deep} utilized pre-trained networks as feature extractors for subsequent classification. 
Byra et al. \cite{byra2020early} fine-tuned the Inception-ResNet-V2 CNN based on the benign and malignant classification tasks and took Siamese architecture for receiving images before and after NAC.
Multi-task learning simultaneously implementing \cite{liu2022early,wu2022integrated} segmentation and treatment response prediction also attracts great attention. 
Liu et al. \cite{liu2022early} proposed a Siamese multi-task network (SMTN) consisting of segmentation sub-networks and pCR predication sub-networks. 
Wu et al. \cite{wu2022integrated} constructed three image signatures based on features extracted from segmentation networks for tumor segmentation of three treatment phases, and then the image signatures were integrated with clinical factors for pCR prediction.

\noindent\textbf{Other Tasks.}
Identifying and classifying the biomarkers or indicators which represents the effect of treatment also play important roles in clinical practice. 
Aghaei et al. \cite{aghaei2015computer} computed kinetic image features and implemented classification of response to chemotherapy based on different fusion combinations of the features with ANN models, and then selected clinical markers according to feature analysis. 
For more specific assessment, some \cite{li2022predicting} developed Stroma-derived bio-marker and then obtained new clinical markers based on CNNs from histological images.

\subsection{Prognosis}

Prognosis aims to evaluate the likely outcome or course of a disease.  
\textcolor{black}{We categorize the applications in this field into classification and other tasks.}

\subsubsection{Classification}

\noindent\textbf{Survival prediction.}

The Cox regression model \cite{breslow1975analysis} is a typical survival prediction method that tries to relate multiple variables to the event of death over time under the proportional hazard assumption. 
Recently, based on the bright field histology images, to feed better features into the Cox regression model for hazard prediction, Morkunas et al. \cite{morkunas2021tumor} proposed to use ANN to segment the collagen fibers and extracted 37 features of collagen fiber morphometry, density, orientation, texture, and fractal characteristics in the entire cohort, and the features were finally analyzed with a cox regression model.
Liu et al. \cite{liu2021tsdlpp} trained a network to find cancerous areas, used a densely connected CNN to extract multi-level image features, and fed the features into the Cox regression model for survival prediction.

Apart from methods using a single modality, Liu et al. \cite{liu2022deep} additionally combined multiple modalities including clinical information (such as sex, age and cancer stage), multi-scale WSIs, as well as the tumor, lymphocyte, and nuclear segmentation results for survival prediction via a MobileNetv2 \cite{Sandler2018MobileNetV2IR}. 
Wang et al. \cite{wang2021gpdbn} combined WSI and genomic features for survival prediction using a bi-linear neural network.

\noindent\textbf{Recurrence prediction.}
Liu et al. \cite{liu2021mask} used a 3D CNN to predict breast cancer recurrence after 5 years as well as the HER2 status based on DCE-MRI. 
Using the immunofluorescence images of CD8+ T lymphocytes and cancer cells, Yu et al. \cite{yu2020predicting} used ``deepflow" from MXNet \cite{Chen2015MXNetAF} for relapse prediction for patients with triple-negative breast cancer.
Ha et al.\cite{ha2019convolutional} proposed to use CNN to predict the Oncotype Dx recurrence score based on MRI images to provide an alternative to the invasive and expensive genetic analysis of Oncotype Dx could be avoided. 
Ma et al. \cite{ma2022radiomics} identified the bio-markers indicating recurrence for TNBC patients with NAC treatments based on radiomics analysis in segmented images by the 3D pre-trained U-nets.

In addition to single-modality methods, multi-modal methods have also been explored in this task. 
Kim et al. \cite{kim2021deep} first identified 32 features related to breast cancer recurrence and developed a recurrent neural network to predict the recurrence time. 
To predict the recurrence and metastasis of HER2 positive breast cancer in patients, Yang et al. \cite{yang2022prediction} proposed to use CNN to extract feature from WSI and combine it with clinical information via a multi-modal model. 
Rabinovici et al. \cite{rabinovici2022multimodal} utilized the ensemble strategy on the prediction scores on parallel CNN-based models for different MRI modalities and clinical classifiers.
To predict the recurrence risk of early-stage breast cancer, Nichols et al. \cite{nichols2020quantitative} used an artificial neural network to combine pathological, clinical, and imaging variables.
Specifically, the global mammographic breast density and local breast density (LBD) are used, and LBD was measured with optical spectral imaging capable of sensing regional concentrations of tissue constituents. This method demonstrated high correlation of risk prediction with Oncotype Dx recurrence score.
Whitney et al. \cite{whitney2018quantitative} used nuclear morphology features from hematoxylin and eosin (H\&E) stained images to predict risks derived by the Oncotype DX test. 
After feature selection, the performance of multiple machine learning methods and a deep neural network are compared.

\subsubsection{Other Task}
The main purpose aside from prognosis prediction is the discovery of prognostic biomarkers \textcolor{black}{and explore the interactions between various prgnostic factors.}

One line of studies uses a neural network as the first step to detect important features. Bai et al. \cite{bai2021open} applied a neural network for the detection of tumor cells, immune cells, fibroblast cells, and others. 
In the end, five machine tumor-infiltrating lymphocyte variables were derived based on features. These variables were found to be independent and robust prognostic indicators.

Another line of research uses neural networks to associate features.
Balkenhol et al. \cite{balkenhol2021optimized} tried to assess the tumour infiltrating lymphocytes (TILs). A CNN was applied to register unmixed multispectral images and corresponding H\&E sections. It was found that for all TILs markers, the presence of a high density of positive cells correlated with improved survival. None of the TILs markers was superior to the others.
Using the graph neural network (GNN), Qiu et al. \cite{qiu2022intratumor} proposed to tape the regional interactions among existing biomarkers (tumor size, nodal status, histologic grade, molecular subtype, etc.) to reveal hidden prognostic values. 
Differently, Lee et al. \cite{lee2022derivation} used a GNN to explore contextual features in gigapixel-sized WSIs in a semi-supervised manner to provide interpretable prognostic biomarkers.

Recently, Zuo et al. \cite{zuo2022identify} combined the WSI data and genomic data to assess the prognostic outcome. This work fused the interaction between WSI and genomic features via the attention mechanism which enabled the identification of survival-associated imaging and genomic biomarkers strongly correlated with the interaction between TILs and tumors.

\subsection{Discussion}

\textcolor{black}{
With the increasing accuracy achieved by deep learning models, more and more studies reported methods that had high performances that are even comparable with radiologists under limited data in mammogram \cite{rodriguez2019stand}, Ultrasound \cite{qian2021prospective}, MRI \cite{witowski2022improving}, and pathology images \cite{bejnordi2017diagnostic}.
Growing numbers of studies are seeking extra clinical applications based on DL.
}

\textcolor{black}{For mammograms, owing to the large-scale data and the need for population-level screening, more and more FDA-cleared or approved DL products are available on the market, such as Lunit INSIGHT MMG, Transpara®, MammoScreen® 2.0, etc.
Several DL algorithms and products could take multi-view inputs and conduct accurate cancer classification and lesion detection at the same time.
There are increasing efforts to predict risks from even normal mammograms \cite{arefan2020deep,yala2021toward,yala2022multi}, which would potentially increase the screening efficacy and effectiveness.
We also noticed that a recent prospective study \cite{ng2023prospective} based on Mia (version 2.0, Kheiron Medical Technologies) showed that using AI as an additional reader can improve the early detection of breast cancer with relevant prognostic features, with minimal to no unnecessary recalls.
Compared with mammograms, ultrasound is more effective in screening for women with dense breasts (which has a high proportion in China) and is more often used for lymph node metastasis estimation and risk prediction.
Despite the prosperity of the studies on breast ultrasound, there is a lack of mature DL products that can be used in clinics.
Also, the scale of public breast ultrasound data is far less than that of mammograms.
}

\textcolor{black}{
Breast MRI is often used preoperatively for treatment planning, which plays a lesser role in cancer screening than mammogram and breast ultrasound. 
This is partially the reason why many of the public breast MRI datasets are aimed at treatment response or prognosis predictions, and studies did show that MRI is effective in outcome prediction \cite{peng2022pretreatment,duanmu2020prediction}.
In addition, current studies also showed that breast MRI could be used to reduce false positives \cite{witowski2022improving} and potentially used in screening \cite{jing2022using,verburg2022deep,bhowmik2023automated}.
Further, it is also of great interest whether the examination process could be more efficient with less time and contrastive agent \cite{ayatollahi2021automatic,chung2022deep}.
}

\textcolor{black}{
DL has demonstrated the capability of analyzing pathology images in various clinical tasks, such as molecular subtyping, mitosis detection, metastasis detection, fine-grained cancer classification, virtual staining, outcome prediction, etc. 
Recent studies are also seeking effective algorithms that can automatically grade the cancers \cite{wang2022improved,wahab_ai-enabled_2023} according to the well-established prognostic factor, Nottingham histological grade (NHG).
A recent promising study by Amgad et al. \cite{amgad2023population} showed a DL-based population-level digital histologic biomarker consistently outperformed pathologists (usually use NHG) in predicting survival outcomes, independent of tumor–node–metastasis stage and pertinent variables.
Besides, it's also quite appealing to find the association between imaging features of pathology images and the genetic information \cite{mondol_hist2rna_2023}.
}

\textcolor{black}{
Outcome prediction, i.e., treatment response prediction and prognosis prediction, is trending in the field of deep learning-based breast cancer analysis.
It usually requires clinicians to summarize a diverse range of different clinical factors and their sophisticated interconnections to predict the future outcome of the patients.
Owing to their remarkable capability of learning patterns from complex high-dimensional data, DL-based breast cancer imaging also shows promise in outcome prediction by providing a non-invasive alternative to conventional analysis.
We found not only an increase of studies in this direction but also the practical values demonstrated by DL techniques whose results could further help clinicians make personalized treatment plans and facilitate the discovery of potential biomarkers \cite{amgad2023population}.
}

\textcolor{black}{To summarize this section, we witnessed the growing scale of research on using deep learning for breast cancer image analysis and the increasing feasibility of DL in assisting real-world clinical applications. In the next section, we will delve into details on the challenges and future directions of DL-based breast cancer imaging from a methodological perspective.}

\section{Challenges and Future Directions}
\label{Challenges and Future Directions}

\subsection{Robust Learning with Limited Data}
Large-scale training data is the key to the success of deep learning.
\textcolor{black}{
There were some efforts to make larger scale data accessible for research purposes \cite{trister2017will,halling2020optimam}.
However, most of the available public datasets are of relatively small scales, especially the modalities other than mammograms as those are not routinely exams (please find our summary of the publicly available datasets in the appendix).
This is a potential factor that hinders the robustness and generalizability of deep learning-based breast cancer analysis models.}
\textcolor{black}{A plausible direction to enlarge the training data while protecting patients' privacy} is federated learning (FL) with the collaboration of multiple institutes.
FL allows jointly training deep learning models without sharing data among participants, which enables cooperation across clients while also preserving the patient's privacy.
\textcolor{black}{Synthetic data is another potential option, and a recent pioneer work provided 2,000 synthetic pathology images for nuclei segmentation studies \cite{ding2023large}.}

Another challenge naturally inherited from limited training data is the limited generalization ability of the developed model, such as the results reported by Wang et al. \cite{wang2020inconsistent}.
Novel algorithms that can improve the robustness of deep learning models on unseen domains (such as domain adaptation and domain generalization) would be of interest in the field of breast cancer analysis \cite{xue2021selective,zhang2019whole}.
Existing approaches are majorly based on learning invariant representations from multi-source data.
A recent study also found that fine-grained annotations could improve a model's generalizability without involving training data from multiple sources \cite{luo2022rethinking}.
It is worth mentioning that domain gaps generally exist under FL scenarios as well, where the data provided by each client are naturally heterogeneous.
In addition, robustness to adversarial attack is also in a need to protect the medical analysis systems from potential threats \cite{zhou2021machine,qi2020stabilized}.

With the development of medical imaging techniques, novel imaging data would also emerge.
Compared with the widely and maturely studied modalities, the new modalities would need further adaptation of current deep learning algorithms.
Transfer learning and domain adaptation hold promises in addressing this challenge, and a typical example is the studies that made efforts to transfer the models learned from mammograms to DBT \cite{samala2016mass}. 

\subsection{Efficient Learning with Weak Labels}
Apart from data, label is another important factor that drives the thriving of deep learning models, and supervised learning is the most common form of deep learning \cite{lecun2015deep}.
For breast cancer imaging, labels are often obtained in two ways: determined by radiologists' interpretation of the radiology images (e.g., BI-RADS 2 as normal, 3 as benign, 4-6 as malignant); or determined by pathological results based on biopsy or surgery, which is also the ``gold standard" for cancer identification.
However, the former strategy inevitably involves inter-reader variation, while the latter is invasive and may not always be available.
Moreover, annotating medical images is labor-exhaustive and expertise-depending, and hence annotations are scarce, especially fine-grained annotations like bounding boxes or segmentation masks.
The medical image annotations often cannot meet the requirement of quantity and quality, and efforts have been made to develop label-efficient learning models to leverage more available data \cite{jin2023label,you2023rethinking}.
Recent studies proposed omni-supervised learning to utilize different types of annotations for training a unified network \cite{luo2021oxnet,chai2022orf}, which could potentially serve as a unified solution to label scarcity.
Considering the inter-reader variation, a potential solution is developing models that are robust to noisy labels or using the calibration of multiple-reader results \cite{ji2021learning,wang2023learning}.
Self-supervised learning could largely mitigate the need for experts' annotations, which would also be a promising solution to label scarcity.
Self-supervised learning pre-trains a network with only unlabeled data has shown remarkable results (e.g., surpassing the performance of ImageNet pre-trained networks) \cite{you2022intra,schirris2022deepsmile}.
Moreover, free-text report can also benefit self-supervised learning, especially in label-efficient finetuning for downstream tasks \cite{zhou2022generalized}.
\textcolor{black}{It is also worth noting that self-supervised learning may require a large amount of data, which is entangled with the challenge of limited data.}

\subsection{Multimodal Learning for Information Fusion}
In this paper, we have surveyed a number of studies with multimodal learning, and most of them involve only one type of imaging data (e.g., multiple views of mammogram; B-mode ultrasound and color Doppler; different sequences of MRI).
However, different imaging techniques provide various insights into the status of breast cancer patients, while current studies have not yet fully utilized the rich context provided by all different modalities generated during the clinical process.
For example, mammograms are more sensitive to calcifications than ultrasounds, MRI provides more detailed spatial information, and pathology images enable observing the microscopic environment and the cell status.
Also, data beyond the images, such as clinical information, molecular biomarkers, genomics, or clinical reports, could further enrich the descriptions for the patients with more structured knowledge and are proved beneficial for developing multimodal learning models \cite{yang2022prediction,qiu2022intratumor}.
\textcolor{black}{For prognosis prediction, multimodal learning is also essential to improve the accuracy, explore the interconnections among the information from different modalities, and facilitate novel biomarkers \cite{xu2023multimodal,zhou2023cross,qiu2022intratumor}.}
Moreover, interconnections among the multimodal data could be further explored.
For instance, the study by Coudray et al. \cite{coudray2018classification} showed that genes could be predicted from pathology images.
The discovery of such associations could narrow down the search space of biomarkers for clinical analysis and possibly provide non-invasive alternatives to biopsy or serology tests as well.

We can witness a trend of increase in multimodal breast cancer papers, and this line of study is yet to be further explored in the near future.
The key factor that hinders this line of work is the difficulty in constructing a dataset where many different modalities such as gene, age, images, therapy, clinical record, etc., for each patient are complete and collected appropriately.
As a result, future multimodal studies may also pay attention to the problem of missing modalities when the provided data are incomplete \cite{qiu2023scratch}.
We look forward to an open-source multimodal breast cancer dataset to stimulate the development of this direction.

\subsection{Reliable and Explainable Model Learning}
For breast cancer, factors such as age and race have long been important factors that affect healthcare disparity \cite{yedjou2019health}, and one of the reasons is that these factors are highly related to breast density which is often evaluated for breast cancer diagnosis.
The statistics by Giaquinto et al. \cite{giaquinto2022breast} also showed that Black women have the lowest survival for every stage of breast cancer diagnosis except for stage I for which survival is similar.
Deep learning unfortunately inherits the unreliable and unfair diagnosis, especially for under-represented groups \cite{seyyed2021underdiagnosis}.
Recent study also found that most of the data supporting approval of AI products by The United States Food and Drug Administration (FDA) did not report the race or ethnicity of the patients \cite{potnis2022artificial}.
As a result, fair deep learning has raised a lot of attention \cite{xu2022survey}.
Another problem raised by group imbalance is shortcut learning \cite{geirhos2020shortcut}, which refers to the phenomenon when DL models learn wrong patterns for making decisions due to spurious correlations.
\textcolor{black}{One of the key reasons for shortcut learning is that the dataset is not diverse enough and the minority group(s) are under-represented and under-learned}.
Shortcut learning is also one of the reasons that DL models perform inconsistently on the training data and external testing data.
Broadly, three types of algorithms could be explored for fair and debias model learning: pre-processing, in-processing, and post-processing, which focus on fair data stratification, fair model development, and fair modification of a trained model's output, respectively.
Most existing solutions often rely on exhaustive labeling of group information, while a recent study has proposed methods without knowing the group information \cite{luo2022pseudo}, which is more feasible when the patients' information is protected.

Explainability is another important factor to achieve reliable and fair deep learning models.
A large proportion of previous works focused on post hoc methods to interpret an already trained model and help the users, i.e., doctors and patients, to understand the decision-making process of the AI models.
A typical example is using saliency maps, such as CAM, to explain which part of an image contributes most to the final prediction.
Global interpretation (e.g., by Shapley Value) that summarizes the holistic decision-making rules based on the whole cohort could also help understand the learned knowledge of the models.
Recently, developing ante hoc algorithms which embed the reasoning process into DL models has gained more attention.
For example, Wang et al. \cite{wang2022knowledge} proposed a prototype-based model which conducts classification by comparing input images with learned prototypes.
Learning and reasoning based on attributes related to diseases is also a promising approach, but may require additional information for labeling the attributes \cite{zhao2021diagnose,yan2023towards}.
These methods could not fully explain the DL models, but have contributed to opening the ``black box" to a certain level.

\subsection{Personalized Treatment Planning}

Optimal therapy for each patient depends on tumor subtype (e.g., HER2 negative, HER2 positive, and triple-negative), anatomic cancer stage, and patient preference \cite{waks2019breast}.
Personalized treatment planning for breast cancer could largely improve patients' life quality by advancing the treatment for patients with good response and avoiding over-treatment for patients with poor response.
In this survey, we have observed an increase of studies on breast imaging-based treatment response prediction and prognosis, of which a considerable proportion utilized a single modality of breast imaging.
Considering the rich context provided by the multimodal information generated during the diagnosis process, it is of great importance to develop multimodal learning algorithms to combine imaging information, medical history, and genetic profile individualized treatment planning. 
Meanwhile, facilitating precise biomarkers is another crucial direction, which enable identifying patients who will benefit from either escalated or de-escalated treatment \cite{goutsouliak2020towards}.
There were works on constructing new biomarkers based on H\&E pathology images \cite{li2021deep,li2022predicting,amgad2023population}, and we look forward to more effective novel biomarkers in the future.

\subsection{Accountable AI Evaluation and Regulation}

With the increasing performance of AI techniques, there are rising calls to establish the accountability standard of deep learning to improve the evidence of its usefulness and fully unleash its huge potential in healthcare.
\textcolor{black}{Currently, many works that reported high performance of DL-based medical image analysis are retrospective studies conducted on limited datasets.}
Future studies should also rethink the evaluation design to improve the strength of evidence for the developed AI systems.
For example, external validation is now required for reporting AI performance in radiology studies \cite{bluemke2020assessing}.
The UK National Screening Committee further suggested that prospective studies should be required to provide further assessment of AI systems in breast screening pathway, as enriched, multi-reader, multi-case, test laboratory studies are also biased \cite{taylor2022uk}.
\textcolor{black}{Generalizability and robustness of the AI decision support tools should be thoroughly evaluated before adoption for patient care in the clinic \cite{hadjiiski2023aapm}, and we've seen some efforts on developing DL systems with more diverse data \cite{xiang2023deep,aubreville2023comprehensive}.
}
Apart from diverse and more transparent evaluation of the developed systems, involving human experts' intervention could also possibly achieve trustable AI.
Human-in-the-loop is a promising approach where doctors can participate in the training of deep learning algorithms by providing the knowledge on the labels or attributes in a medical image to rectify the models' decision-making process \cite{yan2023towards}.

Further, accountability requires stronger and clearer regulation of AI systems.
AI-based software is also emerging with the increasing studies on AI-based breast cancer analysis, and FDA has approved several AI software indicated for breast cancer image diagnosis. 
However, a recent review on FDA-approved or FDA-cleared AI softwares for breast cancer also reported important gaps in validation approaches, based on which the FDA evidentiary regulatory was suggested to be strengthened \cite{potnis2022artificial}.
The efforts by by clinicians, researchers, engineers, ethicists, and the government are needed to collectively ground the AI technology in breast cancer analysis.

\section{Conclusion}
\label{conclusion}

Breast cancer has become the most diagnosed malignancy worldwide, and breast imaging plays a significant role in breast cancer screening, diagnosis, treatment response prediction, and prognosis.
With the ground-breaking development of deep learning research, emerging studies have been conducted to apply deep learning techniques to tackle breast cancer.
To this end, we conducted this survey to review the deep learning-based breast cancer imaging analysis over the past decade.
Specifically, screening and diagnosis has attracted most of the attention from the deep learning community, while the increase of studies for treatment response prediction and prognosis has also been observed.
The findings of this paper suggest that these studies are largely determined by the available data, especially those released to the public.
By discussing the challenges and exploring the potential future directions, we hope to provide novel insights to inspire readers to devote further efforts on developing the next-generation trustworthy healthcare models for breast cancer as well as other diseases.\\

\section{Appendix}
\subsection{Public Datasets}
\label{Datasets and Evaluation Metrics}

We briefly introduce the public datasets currently available for deep learning-based breast cancer image analysis here.
We refer the readers to the related publications and sources for more details about the datasets.
d
\subsection{Mammogram Datasets}

\begin{table*}[ht]
\centering
\begin{tabular}{llp{0.55\textwidth}}
\hline
\textbf{Dataset Name} & \textbf{Number of Images} & \textbf{Annotations} \\ \hline
MIAS  & 322 images from 161 cases & Normal, benign, and malignant. ROIs of lesions \\ 

DDSM & 10,480 images from 2,620 studies & Normal, benign, and malignant. ROI annotations (bounding box). \\

CBIS-DDSM & 10,239 images & Normal, benign, and malignant. ROI annotations (mask). \\

BancoWeb & 1,400 images from 320 patients & Normal, malignant, benign \\

INBreast & 410 images from 115 cases & Asymmetry, calcification, cluster, mass, distortion, spiculated region, pectoral muscle. BI-RADS. Biopsy (for 56 cases).\\

BCDR-FM & 3,703 images from 1,010 cases & BI-RADS, lesion task.\\

BCDR-DM & 3,612 images from 724 patients & BI-RADS, lesion mask.\\

CMMD & 3,728 images from 1,775 patients & benign, malignant. Molecular subtypes (1,498 images).\\

DREAM & $>$640,000 images from $>$86,000 women & cancer, non-cancer\\

OMI-DB & 3,072,878 images from 172,282 women & Normal, malignant, benign. Lesion bounding boxes.\\

VinDr-Mammo & 20,000 images from 5,000 exams & Mass, calcification, asymmetry, distortion, and other associated features, lesion bounding boxes, BI-RADS.\\

BCS-DBT & 22,032 DBT volumes from 3,060 patients & Normal, actionable, biopsy-proven benign, biopsy-proven malignant, lesion bounding boxes (435 cases).\\

\hline

\end{tabular}
\caption{Summary of Public Mammography and DBT Datasets.}
\label{tab:MGDataset}
\end{table*}

\textbf{Mammographic Image Analysis Society (MIAS)\footnote{\url{https://www.repository.cam.ac.uk/handle/1810/250394}} \cite{suckling1994mammographic,suckling2015mammographic}} is a UK database of 322 mediolateral oblique (MLO)-view mammograms from 161 cases, categorized into normal, benign, and malignant.
ROIs of lesions are annotated with circles.
The dataset also has another version, i.e., \textbf{MIAS MiniMammographic (mini-MIAS)} dataset, where the original images are resized and clipped/padded to a fixed 1024$\times$1024 resolution.

\textbf{Digital Database of Screening Mammography (DDSM)\footnote{\url{http://www.eng.usf.edu/cvprg/Mammography/Database.html}} \cite{pub2000digital} } is a US dataset with 2,620 scanned film mammography studies. Each study contains two views, i.e., mediolateral oblique (MLO) view and craniocaudal (CC) view, for each breast, resulting in a total of 10,480 images. 
Cases are labeled as normal, benign, and malignant with pathological verification and manually generated ROI annotations (bounding boxes) for the abnormalities.

\textbf{Curated Breast Imaging Subset of DDSM (CBIS-DDSM)\footnote{\url{https://wiki.cancerimagingarchive.net/pages/viewpage.action?pageId=22516629} } \cite{lee2017curated}} is a subset of the DDSM dataset containing 10,239 images selected and curated by a trained mammographer. Segmentation masks for the lesions are further provided. 

\textbf{BancoWeb\footnote{\url{http://lapimo.sel.eesc.usp.br/bancoweb} } \cite{matheus2011online}} is a Brazil dataset with 1,400 images from 320 patients upon publishing, categorized into malignant, benign, and normal according to previous and later examinations (including follow-up mammograms, ultrasound, and/or biopsy).
 
\textbf{INBreast\cite{moreira2012inbreast}} is a Portugal dataset with 410 mammograms from 115 cases, categorized into asymmetry, calcification, cluster (of MCCs), mass, distortion, spiculated region, and pectoral muscle by specialist.
MLO view and CC view for each breast are available for 90 cases, and the remaining 25 cases have two views of only one breast.
Breast Imaging Reporting and Data System (BI-RADS) from 1 to 6 are provided for each case, and a biopsy was provided for 56 cases.

\textbf{Breast Cancer Digital Repository (BCDR)\footnote{\url{https://bcdr.eu/information/contacts} } \cite{lopez2012bcdr}} is a Portugal dataset which is divided into two subsets: 
(a) \textbf{Film Mammography-based Repository (BCDR-FM)} with 3,703 digitized film mammograms from 1,010 cases, where lesions have biopsy proofs, BI-RADS scores, and segmentation masks;
(b) \textbf{Full Field Digital Mammography-based Repository (BCDR-DM)} with 3,612 mammograms from 724 patients, where lesions have BI-RADS scores and segmentation masks.
\cite{arevalo2016representation} further added 736 images from 344 patients into BCDR-FM.

\textbf{The Chinese Mammography Database (CMMD)\footnote{\url{https://wiki.cancerimagingarchive.net/pages/viewpage.action?pageId=70230508} }} \cite{cui2021chinese} is a Chinese dataset of 3,728 mammograms from 1,775 patients, with biopsy-confirmed benign or malignant tumors. The molecular subtypes are also provided for 739 patients (1,498 mammograms).

\textbf{The Digital Mammography DREAM Challenge }\footnote{\url{https://www.synapse.org/\#!Synapse:syn4224222/wiki/401743}} \cite{trister2017will} provides a US dataset with over 640,000 de-identified mammograms from over 86,000 women, categorized into cancer and non-cancer.
More than 99\% of the examinations have both CC and MLO views for each breast.
The results on this dataset can only be obtained with code uploaded to the challenge.

\textbf{OPTIMAM mammography database (OMI-DB)}\footnote{\url{https://www.cancerresearchhorizons.com/licensing-opportunities/optimam-mammography-image-database-omi-db} }\cite{halling2020optimam} is a UK dataset containing  3,072,878 images from 172,282 women, containing biopsy-proven malignant, benign, and normal cases.
Bounding boxes of the lesions and associated clinical data are also provided. 

\textbf{Breast Cancer Screening – Digital Breast Tomosynthesis (BCS-DBT)\footnote{\url{https://wiki.cancerimagingarchive.net/pages/viewpage.action?pageId=64685580}} \cite{buda2021data}} is a US dataset with 22,032 DBT volumes from 5,060 patients, categorized into normal, actionable (additional imaging was needed but no biopsy was performed), biopsy-proved benign, and biopsy-proved malignant cases.
Further, 435 bounding boxes are provided in the central slice for the biopsy-proved lesions.
Most of the data have four views (MLO and CC views of two breasts).
The complete BCS-DBT dataset is used for the \textbf{DBTex2 Challenge\footnote{\url{https://www.aapm.org/GrandChallenge/DBTex2}}}, while a subset of BCS-DBT with 1,000 DBT scans from 985 patients is used for the \textbf{DBTex Challenge\footnote{\url{https://www.aapm.org/GrandChallenge/DBTex}}}.

\textbf{VinDr-Mammo\footnote{\url{https://vindr.ai/datasets/mammo}} \cite{nguyen2023vindr}} is a Vietnamese dataset of digital mammography with breast-level assessment and extensive lesion-level annotations. The dataset contains 5,000 mammography exams, each with four standard views, resulting in 20,000 images in total. The dataset contains BI-RADS and breast density assessment at the individual breast level. For non-benign findings, the dataset also provides the category (Mass, calcification, asymmetry, distortion, and other associated features), locations (lesion bounding boxes), and BI-RADS.

\subsection{Ultrasound Datasets}

\begin{table*}[h]
\centering
\begin{tabular}{llp{0.45\textwidth}}
\hline
\textbf{Dataset Name} & \textbf{Number of Images} & \textbf{Annotations} \\ 
\hline
BUSI & 780 images from 600 women & Normal, benign, malignant. Lesion Masks.\\

Dataset B/UDIAT & 163 images & Each lesion was delineated by experienced radiologists \\

NCID & $\sim$ 78,000 women & Records are provided. \\

TFUSL & 577 cases & Benign (244), Malignant (130), Pitfall cases (51), Prosthesis (59), Male breast (38), Axilla (26), Elastography (21), 3D imaging (8) \\

STD & 42 images & Lesion masks. \\

TUHD & 180 images from 180 patients & Lesion masks. \\

OABRFD & 100 lesions from 78 women & Malignant, benign. \\

BUSIS & 562 images & Lesion masks. \\

RW & 439 cases & Diagnostic reports and case discussions are provided.\\

BUS-BRA & 1,875 images from 1,064 patients & Biopsy-proven benign and malignant. BI-RADS. Lesion masks.\\

\hline
\end{tabular}
\caption{Summary of Public Breast Ultrasound Datasets.}
\label{tab:US_datasets}
\end{table*}

\textbf{Breast ultrasound images (BUSI) dataset}\footnote{\url{https://scholar.cu.edu.eg/?q=afahmy/pages/dataset}}~\cite{al2020dataset} was collected in 2018 at Baheya Hospital for Early Detection \& Treatment of Women's Cancer, Cairo, Egypt.
The dataset consists of 780 breast ultrasound (BUS) images among 600 women in ages between 25 and 75, with an average image size of $500\times500$ pixels in PNG format. All images were cropped to different sizes to remove unused and unimportant boundaries from the images. The images are categorized into normal, benign, and malignant. The delineation of the breast lesions (lesion mask) is also provided.

\textbf{Dataset B or UDIAT dataset}\footnote{\url{http://www2.docm.mmu.ac.uk/STAFF/m.yap/dataset.php}}~\cite{yap2017automated} was collected in 2012 from the UDIAT Diagnostic Centre of the Parc Taulí Corporation, Sabadell (Spain) with a Siemens ACUSON Sequoia C512 system 17L5 HD linear array transducer (8.5 MHz). The dataset consists of 163 images from different women with a mean image size of $760\times570$ pixels, where 53 images are with cancerous masses and 110 are with benign lesions. Each lesion was delineated by experienced radiologists. 

\textbf{Breast Ultrasonography image datasets at the National Cancer Institute (NCID)}\footnote{\url{https://cdas.cancer.gov/datasets/plco/19/}} provide very comprehensive data that contains nearly all the available data for breast cancer incidence and mortality analyses from The Prostate, Lung, Colorectal and Ovarian (PLCO) Cancer Screening Trial. 
The dataset contains one record for each of the approximately 78,000 women in the PLCO trial.

\textbf{The free ultrasound library by SonoSkills and FUJIFILM Healthcare Europe (TFUSL)}\footnote{\url{https://www.ultrasoundcases.info/cases/breast-and-axilla/}} provides 7,672 cases with
59,336 ultrasound images and clips, covering multiple organs and peripheral vessels. The information about breast and axilla ultrasound data is as follows.
For the regular US imaging, there are 244 benign lesions and 130 malignant lesions. Besides, there are 51 pitfall cases, including 11 malignant lesions mimicking a benign lesion, 16 benign lesions mimicking malignancy, and 24 unusual lesions. For prosthesis, there are 5 normal breast implant cases, 33 ruptured breast implant cases, and 21 miscellaneous breast implant cases. Besides, there are 38 male breast cases and 26 axilla ultrasound images in the dataset. Moreover, elastography and 3D imaging are also included. There are 21 benign and malignant cases for elastography and 8 3D imaging cases. 

\textbf{Shantou Dataset (STD}\footnote{\url{https://github.com/xbhlk/STU-Hospital}}~\cite{zhuang2019rdau} was collected at the Imaging Department of the First Affiliated Hospital of Shantou, consisting of 42 beast ultrasound images. The images were acquired using the GE Voluson E10 Ultrasound Diagnostic System (L11-5 50mm broadband linear array transducer, 7.5MHz frequency) and manually segmented and labeled by the specialist with more than 7-year of experience at the First Affiliated Hospital of Shantou University.

\textbf{Thammasat University Hospital dataset (TUHD) }\footnote{\url{http://www.onlinemedicalimages.com/index.php/en/}}~\cite{rodtook2018automatic} was collected by Thammasat University Hospital of Thailand. The dataset consists of 60 US images of breast cancer, 60 images of cysts,  and 60 images of fibroadenoma from 180 different patients, which were obtained by a Philips iU22 ultrasound machine. The ground truth contours have been hand-drawn by three leading experts from the Department of Radiology of Thammasat University using an electronic pen and a Samsung Galaxy Tablet computer. The final ground truth was obtained by majority vote (two out of three). The image resolutions range from $200\times200$ to $300\times400$ pixels.

\textbf{Open access breast RF database (OABRFD)}\footnote{\url{https://zenodo.org/records/545928}}\cite{piotrzkowska2017open} includes ultrasonic radio-frequency echoes that were recorded from breast focal lesions of patients from the Institute of Oncology in Warsaw. 
Patients were examined by a radiologist with 18-year experience in ultrasonic examination of breast lesions. The set of data includes scans from 52 malignant and 48 benign breast lesions recorded in a group of 78 women. For each lesion, two individual orthogonal scans from the pathological region were acquired with the Ultrasonix SonixTouch Research ultrasound scanner using the L14-5/38 linear array transducer. All malignant lesions were histologically confirmed by core needle biopsy. In the case of benign lesions, a part of them were histologically assessed and the rest were observed over a 2-year period.

\textbf{Benchmark for Breast Ultrasound Image Segmentation (BUSIS) dataset}\footnote{\url{http://cvprip.cs.usu.edu/busbench/}}\cite{zhang2022busis} is composed of 562 B-mode BUS images among women between the ages of 26 to 78. The images were originally collected and de-identified by the Second Affiliated Hospital of Harbin Medical University, the Affiliated Hospital of Qingdao University, and the Second Hospital of Hebei Medical University, using multiple ultrasound devices including GE VIVID 7 and LOGIQ E9, Hitachi EUB-6500, Philips iU22, and Siemens ACUSON S2000. 
Four experienced radiologists were involved in the ground truth generation: three radiologists read each image and delineate each tumor boundary individually, and the fourth one (a senior expert) determines if the majority voting results need adjustments. Hence, the binary mask of lesions are provided for each image.

\textbf{Radiopaedia website (RW)}\footnote{\url{https://radiopaedia.org/}} provides a dataset containing 439 breast tumor ultrasound images. Diagnostic reports and corresponding case discussions are available for each case.

\textbf{BUS-BRA}\footnote{\url{https://zenodo.org/records/8231412}}\cite{gomez2023bus} contains 1,875 images from 1,064 patients who underwent routine breast studies. The dataset includes biopsy-proven tumor cases and BI-RADS annotations in categories 2, 3, 4, and 5 as well as ground truth delineations that divide the BUS images into tumoral and normal regions.

\subsection{MRI Datasets}

\begin{table*}[h]
\centering
\begin{tabular}{llp{0.45\textwidth}}
\hline
\textbf{Dataset Name} & \textbf{Number of Images} & \textbf{Annotations} \\ \hline

Duke & 922 patients & Locations of lesions, demographic, clinical, pathology, treatment, outcomes, and genomic data \\ 

QIN-Breast & 68 patients & Treatment response in the neoadjuvant setting. \\

ISPY1 & 222 subject & Response to treatment and risk-of-recurrence in patients with stage 2 or 3 breast cancer receiving neoadjuvant chemotherapy (NACT). Lesion mask. \\

ISPY2 & 719 patients & Efficacy of new agents for breast cancer in neoadjuvant chemotherapy (NAC) setting. Clinical Data.\\

ACRIN-6698 & 277 women & NAC response. Clinical Data.\\

Breast-MRI-NACT-Pilot & 64 subjects & Treatment response to neoadjuvant chemotherapy (NACT). Lesion Masks. Clinical data.\\

TCGA-BRCA & 164 studies from 139 patients & Genomics, biomedical data, and clinical data are provided.\\

TCGA-Breast-Radiogenomics & 84 patients & The same as TCGA-BRCA.\\

BREAST-DIAGNOSIS & 148 studies from 88 subjects & Pathology results, pathology report, and MRI report. \\

\hline
\end{tabular}
\caption{Summary of the Breast Cancer MRI dataset}
\label{tab:mri_dataset}
\end{table*}

\textbf{Duke Breast Cancer MRI (Duke)}\footnote{\url{https://wiki.cancerimagingarchive.net/pages/viewpage.action?pageId=70226903} } \cite{saha2018machine} collected 922 biopsy confirmed invasive breast cancer patients with breast cancer from the retrospective study of decade years. 
Each subject contains a non-fat saturated T1-weighted sequence, a fat-saturated T1-weighted pre-contrast sequence, and mostly three to four post-contrast sequences.
The dataset also provides non-imaging information such as demographics, treatments, tumor characteristics, recurrence, etc., which could help researchers implement multiple further tasks. 

\textbf{QIN-Breast}\footnote{\url{https://wiki.cancerimagingarchive.net/display/Public/QIN-Breast} } \cite{li2015multiparametric} contains two different modalities: longitudinal PET/CT and quantitative MR which covers 68 patients captured at three different time points: the start of treatment (t1), after the first cycle of treatment (t2), and either after the second cycle of treatment or at the completion of all treatments (prior to surgery) (t3). The MRI data includes diffusion-weighted images (DWIs), DCE-MRI, and multi-flip data for T1-mapping.
Also, patient-level labels for treatment response (pCR/non-pCR) have been assigned for monitoring the response to therapies.

\textbf{I-SPY 1/ACRIN 6657 trials (ISPY1)}\footnote{\url{https://wiki.cancerimagingarchive.net/pages/viewpage.action?pageId=20643859}} \cite{hylton2016neoadjuvant} contains 222 subject. The image acquisition protocol included a localization scan and T2-weighted sequence followed by a contrast-enhanced T1-weighted series.
MRI exams were performed at four different time points to evaluate the treatment response of patients as well as risk-of-recurrence. 

\textbf{I-SPY 2 Breast Dynamic Contrast Enhanced MRI (ISPY2)}\footnote{\url{https://wiki.cancerimagingarchive.net/pages/viewpage.action?pageId=70230072} }\cite{li2020predicting} currently contains 719 patients from three collections.
The first collection contains 719 patients with locally advanced breast cancer who has received neoadjuvant chemotherapy. The MRI scans were acquired from 22 different clinical centers and each patient underwent 4 MRI exams before and after the treatment. Except for DCE-MRI and T2-weighted MRI, it also contains the histopathologic and demographic data such as race, ethnicity, age, drug arms menopausal status, Receptor subtypes, and pCR results. 
The dataset is aimed for evaluating the efficacy of new agents for breast cancer in neoadjuvant chemotherapy (NAC)

\textbf{ACRIN 6698/I-SPY2 Breast DWI (ACRIN-6698)}\footnote{\url{https://wiki.cancerimagingarchive.net/pages/viewpage.action?pageId=50135447}} enrolled 406 women, of which 277 were randomized to experimental treatment or control arms.
DWI, T2-weighted, and DCE MRI were provided. It aims for assessing breast cancer response to neoadjuvant chemotherapy (NAC).

\textbf{Single site breast DCE-MRI data and segmentations from patients undergoing neoadjuvant chemotherapy (Breast-MRI-NACT-Pilot)}\footnote{\url{https://wiki.cancerimagingarchive.net/pages/viewpage.action?pageId=22513764} } contains 64 subjects with invasive breast cancer who are undergoing neoadjuvant chemotherapy. The MRI scans contain the longitudinal DCE-MRI examinations obtained before NAC, after one cycle of NAC and before the surgery. The dataset also contains clinical information and lesion segmentation masks for further research. This trial assesses the Recurrence-free survival (RFS) at 6-month or 1-year intervals after surgery and other demographic and clinical information (e.g., tumor size, histologic type, subtype, and lymph node involvement) have also been provided.  

\textbf{The Cancer Genome Atlas Breast Invasive Carcinoma Collection (TCGA-BRCA)}\footnote{\url{https://wiki.cancerimagingarchive.net/pages/viewpage.action?pageId=3539225}} collected 164 studies from 139 patients with breast cancer. It includes two kinds of modalities: MRI and mammograms. Other corresponding information is also provided as follows: tissue slide images; clinical data that describes lesion malignancy types, pathological stage, and molecular subtypes, corresponding therapy records containing pharmaceutical therapy, radiation therapy, clinical stage, and so on; biomedical data that describes tumors information; demographics; genomics. 
The dataset is aimed at connecting cancer phenotypes to genotypes.

\textbf{TCGA-Breast-Radiogenomics}\footnote{\url{https://wiki.cancerimagingarchive.net/pages/viewpage.action?pageId=19039112}} is filtered from TCGA-BRCA for minimizing the variance and thus achieving more precise performance. This collection contains 84 patients from 4 medical centers. It also contains the annotations of lesion areas, multi-gene assays, and clinical data describing molecular types, tumor stage, lymph node involvement, etc. 

\textbf{BREAST-DIAGNOSIS}\footnote{\url{https://wiki.cancerimagingarchive.net/display/Public/BREAST-DIAGNOSIS}} collected 148 studies from 88 subjects containing breast lesions over high-risk normals, Ductal carcinoma in situ (DCIS), fibroids and lobular carcinomas. It collected data across four different modalities: MRI, mammogram, CT, and PT. The dataset also includes clinical and pathological information which describing the tumor pathological types, molecular subtypes, BI-RADS grades, MRI features, and pathological reports from physicians. 

\subsection{Pathology Image Datasets}

\begin{table*}[h]
\centering
\begin{tabular}{cp{0.3\textwidth}p{0.48\textwidth}}
\hline
\textbf{Dataset Name} & \textbf{Number of Images} & \textbf{Annotations} \\ \hline

MITOS & 5 H\&E WSIs &  Location of mitosis.\\ 

MITOS-ATYPIA-14 & 1420 H\&E frames &  Location of mitosis. Nuclear atypia score.\\ 

CAMELYON16 & 400 H\&E WSIs & WSI binary masks for metastasis.\\

CAMELYON17 & 1,000 H\&E WSIs & pN-stage of a patient. Metastases lesion location (for 50 slides) \\

TUPAC16* & 821 WSIs & Mitotic scores. \\

BACH & 500 microscopy images, 40 WSIs & Normal, Benign, in sity carcinoma, invasive carcinoma (microscopy image). Binary WSI mask (the same four classes). \\

Breastpleomorphism & 118 H\&E WSI &  Nuclear pleomorphism
scoring \\

BreCaHAD & 162 H\&E images & mitosis, apoptosis, tumor nuclei, non-tumor nuclei, tubule, and non-tubule.\\

HRE2C & 172 WSIs from 86 cases stained with both H\&E and IHC & HER2 scores. \\

BreaKHis & 7,909 microscopy images from 82 patients & Benign and malignant. \\

MIDOG-2021 & 300 cases & Mitotic figures \\

MIDOG-2022 & 520 cases & Mitotic figures \\

UCSB Bio-Segmentation & 58 H\&E images & Malignant, benign. \\

ABCTB & 2535 H\&E images from 2535 patients & Associated bioinformation. \\

MoNuSeg & 44 H\&E images & Nuclear boundary. \\

TNBC & 50 images & Nuclei annotation. Omics information.\\

ISL & 135 microscopy images (2 are breast cancer) &  fluorescent labels \\

BCData & 1,338 Ki-67-stained images from 394 cases & Positive or negative tumor cell. Centroid annotations. \\

HASHI & H\&E slides from near 500 cases & Breast cancer region mask. \\

BreastPathQ & 3,698 patches extracted from 96 H\&E WSI from 55 patients & Cellularity score. \\

TCGA$\S$ & 1,133 H\&E WSI from 1,062 cases & Tumor type, associated clinical data, genomic data, etc. \\

SNOW$\dagger$ & 2,000 images & Nuclei mask.\\

MIDOG++ & 11,937 mitotic figures from 503 histological specimens &  breast carcinoma, lung carcinoma, lymphosarcoma, neuroendocrine tumor, cutaneous mast cell tumor, cutaneous melanoma, and (sub)cutaneous soft tissue sarcoma. \\

\hline
\end{tabular}
\caption{Summary of the Breast Cancer MRI dataset. *: We report the main dataset from TUPAC16, and more details can be found in the supplemenraty text. $\dagger$: SNOW is a synthetic dataset. $\S$: We only report the breast cancer pathology images from TCGA here.}
\label{tab:path_dataset}
\end{table*}

\textbf{Mitosis Detection in Breast Cancer Histological Images (MITOS)
\footnote{\url{https://ludo17.free.fr/mitos_2012/dataset.html}}}~\cite{ludovic2013mitosis} provides the pathology images for mitosis detection in breast tissue. 
The ground truth was provided by expert pathologists for all mitotic cells. The first version contains 5 H\&E-stained Whole Slide Images (WSIs) with 10 annotated microscope high power fields per slide. 

\textbf{ICPR MITOS-ATYPIA challenge 2014 (MITOS-ATYPIA-14)\footnote{\url{https://mitos-atypia-14.grand-challenge.org/Dataset}}} was made up of two parts: detection of mitosis and evaluation of nuclear atypia score.
It contains in total 284 frames extracted from WSIs of 20$\times$ magnification and 1,136 frames from WSIs of 40$\times$ magnification.
The slides were stained with standard hematoxylin and eosin (H\&E) dyes.
Two pathologists annotated nuclei as true mitosis, probably a mitosis, or not a mitosis.
Three junior pathologists categorize nuclear atypia as low, moderate, and high grade, respectively.

\textbf{CAMELYON16\footnote{\url{https://camelyon16.grand-challenge.org}}}~\cite{bejnordi2017diagnostic} was a challenge for detection of metastases in H\&E WSIs of lymph node sections from breast cancer patients.
The data were collected from the Radboud University Medical Center and the University Medical Center Utrecht.
The dataset contains a total of 400 WSIs and an equivalent amount of masks defining the region of metastatic.

\textbf{CAMELYON17\footnote{\url{https://camelyon17.grand-challenge.org}}}~\cite{bandi2018detection} followed the former CAMELYON16 challenge while focusing on patient-level analysis instead of slide-level analysis.
The dataset contains 1,000 WSIs of sentinel lymph nodes with 5 slides per patient.
The pN-stage (which determines whether the cancer has spread to the regional lymph nodes based on the tumor, node, metastasis (TNM) staging system~\cite{tio1996tnm}) annotations of the patient are given. 
There are also 50 slides annotated at the lesion level to locate metastases.

\textbf{TUmor Proliferation Assessment Challenge 2016 (TUPAC16)\footnote{\url{https://tupac.grand-challenge.org/}}}~\cite{veta2019predicting} was a challenge for predicting the tumor proliferation scores from the WSIs. 
The dataset contains in total of 500 WSIs for training and 321 WSIs for testing.
The study also provides two additional datasets to develop a mitosis detection algorithm: a dataset with annotated mitotic figures (73 cases) and a dataset with annotations for regions of interest (148 cases).

\textbf{BreAst Cancer Histology (BACH)\footnote{\url{https://iciar2018-challenge.grand-challenge.org}}}~\cite{aresta2019bach} is a challenge dataset with two goals: 1) automatically classifying H\&E-stained breast histology microscopy images in four classes, i.e., normal, benign, in situ carcinoma, and invasive carcinoma; and 2) segmentation on the microscopy images for the same four classes.
The dataset contains in total 400 microscopy images for training and 100 for testing.
It also contains 40 WSI, in which 30 are used for training and 10 are for testing.

\textbf{Breastpleomorphism\footnote{\url{https://breastpleomorphism.grand-challenge.org/}}} \cite{mercan2022deep} contains 118 H\&E-stained WSIs of breast cancer surgical resections at Radboud University Medical Centers, Nijmegen (The Netherlands), labeled by an international panel of 10 pathologists for nuclear pleomorphism scoring.

\textbf{BreCaHAD\footnote{\url{https://figshare.com/articles/dataset/BreCaHAD_A_Dataset_for_Breast_Cancer_Histopathological_Annotation_and_Diagnosis/7379186}}}~\cite{aksac2019brecahad}, which contains 162 breast cancer H\&E-stained images (1360$\times$1204 pixels) of six classes: mitosis, apoptosis, tumor, non-tumor, tubule, and non-tubule.

\textbf{HER2 Challenge Contest (HRE2C) \footnote{\url{https://warwick.ac.uk/fac/cross_fac/tia/data/her2contest/}}}\cite{qaiser2018her} provides 172 WSIs from 86 cases stained with both H\&E and IHC for HER2 scoring (consensus score from at least twoexpert).

\textbf{The breast cancer histopathological image classification dataset (BreaKHis)\footnote{\url{https://web.inf.ufpr.br/vri/databases/breast-cancer-histopathological-database-breakhis/}}}~\cite{spanhol2015dataset}, which is composed of 7,909 microscopic images (2,480 benign, 5,429 malignant) collected from 82 patients (24 benign, 58 malignant) at P\&D laboratory in Brazil. 

\textbf{MItosis DOmain Generalization Challenge 2021 (MIDOG 2021)\footnote{\url{https://imig.science/midog2021/the-dataset/}}}~\cite{aubreville2023mitosis} contains a training set of 200 cases, split across four scanning systems. 
The training set contains 1,721 mitotic figures (MF) and 2,714 hard examples (non-mitotic figures).
It also contains a test set with an additional 100 cases split across four scanning systems.
The dataset is aimed for mitotic figures detection.

\textbf{MItosis DOmain Generalization Challenge 2022 (MIDOG 2022)\footnote{\url{https://imig.science/midog/the-dataset}}}. The training dataset consists of regions of interest (ROIs) selected by an experienced pathologist from a selection of tumor types (e.g., canine lung cancer, human breast cancer, etc.).
The training set contains 9,501 mitotic figures (MF) and 11,051 hard examples (non-mitotic figures).
It also contains a test set of 100 independent tumor cases and a preliminary test set of 20 cases.

\textbf{UCSB Bio-Segmentation \footnote{\url{https://bioimage.ucsb.edu/research/bio-segmentation}}}~\cite{gelasca2008evaluation} contains 58 H\&E-stained histopathology images, in which 32 images are benign and 26 images are malignant.
Each image is labeled with pixel-level masks of cell nucleus.

\textbf{The Australian Breast Cancer Tissue Bank (ABCTB)\footnote{\url{https://www.abctb.org.au/abctbNew2/default.aspx}}}~\cite{carpenter2014australian} contains biospecimens from 6,225 donors as at 15 May 2014. The first collection centre for the ABCTB was founded at Westmead with the first donor being recruited in 2006. 
Associated information is provided including but not limited to clinical characteristics, disease status, clinical data, longitudinal data, etc.
From this dataset, 2,535 H\&E images from 2,535 patients have been used in a breast cancer hormonal receptor status determination study \cite{naik2020deep}.

\textbf{Multi-Organ Nucleus Segmentation (MoNuSeg)\footnote{\url{https://monuseg.grand-challenge.org/}}}~\cite{kumar2017dataset}. The training dataset consists of 30 H\&E-stained tissue images captured at 40$\times$ magnification, with around 22,000 nuclear boundary annotations. The test set consists of 14 H\&E-stained images with 7000 nuclear boundary annotations. Note that the images are collected from multiple organs, including breast, kidney, liver, etc.

\textbf{Triple Negative Breast Cancer (TNBC)\footnote{\url{https://rgcb.res.in/tnbcdb/}}}~\cite{raju2014triple} consists of annotated H\&E-stained histology images at 40$\times$ magnification. All slides were taken from a cohort of Triple Negative Breast Cancer patients and were scanned with Philips Ultra Fast Scanner 1.6RA. In total, the dataset consists of 50 images with a total of 4,022 annotated nuclei.
The database provides omics information for the cancer tissues.

\textbf{ISL\footnote{\url{https://github.com/google/in-silico-labeling}}}~\cite{christiansen2018silico} contains 135 high-resolution microscopy images from five different laboratories with different transmitted lights, where two of the images are human breast cancer line. For each transmitted light image, 13 2D images are extracted across the z-depth.

\textbf{BCData\footnote{\url{https://sites.google.com/view/bcdataset}}}~\cite{huang2020bcdata} contains 1,338 Ki-67-stained images with 181,074 annotated cells (centroid annotations) from 394 cases belonging to positive or negative tumor cells, created from a consensus of ten pathologists. 

\textbf{HASHI dataset\footnote{\url{https://datadryad.org/stash/dataset/doi:10.5061/dryad.1g2nt41}}}~\cite{cruz2018high} contained H\&E-stained histological slides of 40$\times$ magnification from patients with ER+ breast cancer from near 500 cases. Masks of breast cancer regions are provided.

\textbf{The Cancer Genome Atlas (TCGA)\footnote{\url{https://www.cancer.gov/about-nci/organization/ccg/research/structural-genomics/tcga}}}~\cite{cancer2012comprehensive} is a landmark cancer genomics program, molecularly characterized over 20,000 primary cancer and matched normal samples spanning 33 cancer types. 
There are 1,133 whole slide images (from 1,062 cases with breast cancer) in the TCGA database.
The types of cancers are given, and other information such as clinical data, genomics, etc., are also provided.

\textbf{BreastPathQ}\footnote{\url{https://breastpathq.grand-challenge.org/}}~\cite{petrick2021spie} is a challenge dataset for cancer cellularity scoring. The training, validation, and testing sets contain 2,394, 185, and 1,119 patches extracted from 96 H\&E WSI from 55 patients. This dataset is aimed for cancer cellularity scoring in breast cancer pathology images following neoadjuvant treatment.

\textbf{Synthetic Nuclei and annOtation Wizard (SNOW)}\footnote{\url{https://zenodo.org/records/6633721}}\cite{ding2023large} is a synthetic pathological image dataset. 
By using off-the-shelf image generator and nuclei annotator, the dataset provides 2,000 image tiles and 1,448,552 annotated nuclei.
The aim of this dataset is to develop nuclei segmentation models in a more cost-effective manner.

\textbf{MIDOG++}\footnote{\url{https://github.com/DeepMicroscopy/MIDOGpp}}\cite{aubreville2023comprehensive} is an extension of the MIDOG 2021 and MIDOG 2022 datasets, containing 11,937 mitotic figures from 503 histological specimens.
The labels include breast carcinoma, lung carcinoma, lymphosarcoma, neuroendocrine tumor, cutaneous mast cell tumor, cutaneous melanoma, and (sub)cutaneous soft tissue sarcoma. 
It is worth mentioning that the specimens were from several laboratories with diverse scanners.

\subsection{Full Tables of Surveyed Papers}
We reported here the lists of the papers surveyed in this paper up till December 2022.

\begin{table*}[t!]
	\centering
        \scriptsize
         \caption{Overview of deep learning-based mammogram screening and diagnosis. CL=Classification, SE=Segmentation, DE=Detection, OT = Other tasks.}
 	\label{tab:screening_and_diagnosis_mammogram}
	\begin{tabularx}{\textwidth}{p{0.06\textwidth}p{0.05\textwidth}p{0.8\textwidth}<{\raggedright}}
		\toprule
		\textbf{Reference} & \textbf{Task(s)} &  \textbf{Description} \\
		\midrule

        \cite{suganthi2012improved} & CL     & Feature extraction and selection followed by ANN for breast cancer classification.\\

        \cite{liu2013multitarget}    & CL & ANN with extracted features for breast cancer classification.\\ 

        \cite{dheeba2014computer}  & CL    & Particle Swarm Optimized Wavelet Neural Network for breast cancer classification.\\

        \cite{carneiro2015unregistered} & CL & Multi-view input with several transfer learning steps for breast cancer classification.  \\

        \cite{fonseca2015automatic}  & CL & 3-layer CNN for breat density classification.\\

        \cite{PENG2016134}  & CL & ANN + extracted features for breast cancer classification.\\

        \cite{dhungel2016automated}  & CL & CNN + random forest for breast cancer classification.\\

        \cite{arevalo2016representation} & CL & CNN for breast cancer classification.\\
        
        \cite{sun2017enhancing}  & CL & Graph-based semi-supervised classification with hand-crafted features + CNN.\\

        \cite{teare2017malignancy} & CL & Inception-V3 pre-trained on ImageNet for malignancy classification.\\

        \cite{carneiro2017automated}    & CL & Transfer Learning + multi-view CNN for breast cancer classification.\\

        \cite{zhu2017deep}   & CL & Pre-trained AlexNet + sparse multi-instance learning + ensemble for breast cancer classification.\\
        
        \cite{antropova2017deep} & CL & Multi-level features from a pre-trained CNN + handcrafted features + SVM classifiers for benign and malignant classification. \\
        
        \cite{li2017deep} & CL & Pre-trained AlexNet + radiographic texture analysis for risk assessment. \\
        
        \cite{dhungel2017fully}  & CL & Multi-view network with residual connection structure.\\
        
        \cite{sert2017ensemble} & CL & GoogLeNet + data augmentation + model ensemble for breast cancer classification. \\

        \cite{dgani2018training}   & CL & Noisy channel to train noise-robust CNN for breast cancer classification.\\

        \cite{mohamed2018deep}   & CL & Modified AlexNet with pre-trained weights for breast density classification.\\

        \cite{yousefi2018mass}   & CL & 2D CNN with muliple-instance random forest summarizing 3D prediction for breast cancer classification.\\

        \cite{shams2018deep}   & CL & CNN + feature enhancing with auxiliary generative adversarial training for breast cancer classification.\\

        \cite{samala2018breast}   & CL & Multi-stage transfer learning for breast cancer classification from DBT.\\

        \cite{akselrod2019predicting}   & CL & CNN + XGBoost for breast cancer classification. \cite{chen2016xgboost}\\ 

        \cite{li2019signed}   & CL & Signed graph regularized DNN + adversarial generative data augmentation for breast cancer classification.\\

        \cite{perek2019learning}   & CL & LSTM-based fusing of prior image and current image features for breast cancer classification.\\

        \cite{shen2019globally} & CL & CNN + attention + multiple instance learning for breast cancer classification.\\
        
        \cite{rodriguez2019stand} & CL & Custom CNN compared with 101 radiologists.\\ 
        
        \cite{rodriguez2019detection} & CL & Custom CNN to improve radiologist in finding cancers from mammograms.\\ 

        \cite{rodriguez2019can}   & CL & Custom CNN to determine high risk mammograms for radiologists' double check.\\ 

        \cite{wu2019deep} & CL & Multi-view ResNet-based CNN + pre-training for breast cancer classification. \\ 

        \cite{tardy2019uncertainty}    & CL & Outlier rejection with uncertainty measurement for breast cancer classification.\\

        \cite{li2020digital}  & CL & VGG and custom CNN to compare CNN-based mammogram and DBT-based classification. \\ 

        \cite{wang2020inconsistent} & CL & Multi-center validation to show CNN-based breast cancer classification performance drop on external data. \\

        \cite{clancy2020deep} & CL & Comparative study on effectiveness of different pre-training stratgies. \\

        \cite{shu2020deep}   & CL & DenseNet169 + multiple instance learning for breast cancer classification.\\

        \cite{yan2020multi}  & CL & YOLO + multi-view inputs + multi-task learning for mass classification.\\ 

        \cite{cogan2020deep}  & CL & Multi-view inputs + MobieNetV2 \cite{sandler2018mobilenetv2} + feature fusion for breast density classification.\\

        \cite{mckinney2020international}    & CL & Ensemble of three multi-view custom CNNs for breast cancer screening. \\

        \cite{roth2020federated}    & CL & DenseNet121 with federated learning for breast density classification. \\

        \cite{lotter2021robust}   & CL & ResNet50 + RetinaNet + weakly-supervised multiple instance learning for breast cancer classification. \\

        \cite{ragab2021framework}  & CL & Feature fusion of different CNNs +  support vector machine for breast cancer classification.\\ 
        
        \cite{rv2021augmenting}   & CL & CNN + multiple-level transfer learning for breast cancer classification.\\

        \cite{nebbia2021radiomics}   & CL & CNN + radiomics feature-derived weighting loss + curriculum learning for breast cancer classification.\\

        \cite{cao2021supervised} & CL & Contrastive pre-training learning + multi-view inputs fine-tuning for breast cancer classification.\\

        \cite{tardy2021trainable}  & CL & CNN + multiple instance learning + attention mechanism for breast cancer classification.\\

        \cite{yala2021toward} & CL & ResNet18 + transformer to predict additive risk each year for survival time prediction\\

        \cite{wang2021dae}  & CL & Multi-task CNN: reconstruction + attribute learning +  breast cancer classification.\\ 

        \cite{bai2022feature}    & CL & Multiple inputs (current and prior pairs) + feature fusion for breast cancer classification.\\

        \cite{lee2021identifying}    & CL & CNN for breast cancer classification on conditional GAN-simulated samples. \\

        \cite{laang2021identifying}    & CL & A set of image classifiers and CNNs for suspicous lesion detection + another set of CNNs for risk estimation. \\ 

        \cite{zhou2021machine}  & CL & Developing CNN robust to adversarial attack. \\

        \cite{li2021domain}   & CL & Multi-view input + multiple contrastive learning to improve generalization of CNN on external data.\\

        \cite{wimmer2021multi}    & CL & Task-specific multi-view models for breast density classification, cancer classification and detection + patient-level fusion for cancer classification.\\
        
        \cite{chen2022multi}    & CL & Multi-view inputs + local co-occurrence learning + global consistency learning for breast cancer classification.\\
        
        \cite{wang2022knowledge}    & CL & Prototype learning + knowledge distillation for breast cancer classification.\\
                
        \cite{mohamed2022bilateral}  & CL & Multi-view inputs + siamese CNN for breast risk classification.\\

        \cite{nguyen2022novel}    & CL & Multi-view input + EfficientNet \cite{tan2019efficientnet} + two light GBM \cite{ke2017lightgbm} branches for BI-RADS and density classification, respectively. \\

        \cite{you2022intra}  & CL & Multi-view inputs+ multi-task learning with classification and contrastive learning.\\

        \cite{hsu2022external}   & CL & Ensemble with 11 highest-performing AI models from the DREAM challenge for external validation.\\
        
        \cite{leong2022microcalcification} & CL & Image enhancement + different classification networks for microcalcification classification.\\

        \cite{adedigba2022performance}  & CL & Comparative study of different classification networks on small dataset.\\

        \cite{lehman2022deep} & CL & Comparison of deep learning model and traditional methods on breast cancer risk prediction.\\

        \cite{ayana2022patchless}  & CL & EfficientNet with multi-stage transfer learning for mass classification. \\

        \cite{oyelade2022novel}  & CL & Wavelet transformation + GAN-based data augmentation + wavelet CNN for breast cancer classification.\\

        \cite{samala2014digital}  & DE & Hand-crafted lesion candidate + CNN for false positive resudction.\\

        \cite{samala2016mass}  & DE & Hand-crafted lesion candidate + CNN with transfer learning for microcalcification clusters detection.\\

        \cite{liu2019unilateral} & DE & Bilateral image input + coarse alignment + Mask-RCNN + bilateral contrast module for mass detection. \\

        \cite{shen2020multicontext} & DE & FCN for mass candidates extraction + multi-context multi-task CNN for false positive reduction.\\

        \cite{agarwal2020deep}   & DE & Faster-RCNN for mass detection.\\

        \cite{yang2020momminet} & DE & Fusion of multi-view networks for mass detection.\\ 
        
        \cite{liang2020weakly} & DE & Weakly-supervised detector trained with CAM-generated pseudo labels for breast cancer detection.\\

        \multicolumn{3}{r}{\footnotesize(\textit{continued on the next page})}\\
	\end{tabularx}
\end{table*}

\begin{table*}[t!]
\ContinuedFloat
	\centering
        \scriptsize
        \caption{Overview of deep learning-based mammogram screening and diagnosis. CL=Classification, SE=Segmentation, DE=Detection.}
	\begin{tabularx}{\textwidth}{p{0.06\textwidth}p{0.08\textwidth}p{0.8\textwidth}<{\raggedright}}
		\toprule
		\textbf{Reference} & \textbf{Task(s)} &  \textbf{Description} \\
		\midrule
        
        \cite{yang2021momminet}  & DE & Fusing of multi-view networks improved from \cite{yang2020momminet} for mass detection.\\

        \cite{buda2021data} & DE & YOLO based on DenseNet for mass and architectural distortion detection.\\
        
        \cite{liu2021compare}    & DE & Multi-view inputs with improved Mask-RCNN for soft-tissue lesion detection.\\

        \cite{liu2020cross}    & DE & Multi-view (ipsilateral views) inputs + graph convolutional network-based reasoning for mass detection.\\

        \cite{liu2021act}    & DE & Multi-view inputs + graph convolutional network-based reasoning for mass detection.\\
        
        \cite{hossain2022developing} & DE & Ensemble with modified YOLO-v5 + dataset augmentation for breast lesion detection.\\

        \cite{zhao2022check}    & DE & Multi-view transformer-based detector for mass detection. \\

        \cite{li2022architectural}    & DE & 2D candidate detection + 3D aggregation + false positive reduction for architectural distortion detection.\\

        \cite{zhu2018adversarial}    & SE & Multi-scale FCN + CRF + adversarial training for mass segmentation.\\

        \cite{singh2018conditional}    & SE & Conditional GAN \cite{isola2017image} for mass segmentation.\\

        \cite{abdelhafiz2020convolutional}    & SE & Modified U-Net for mass segmentation.\\

        \cite{li2021sap}  & SE & Improved conditional GAN + superpixel average pooling for mass segmentation.\\
        
        \cite{yan2021two}    & SE & Localizing mass with modified YOLOv3, then segmenting with modified U-Net++ for mass segmentation.\\

        \cite{kushibar2022layer}    & SE & Modified U-Net + uncertainty estimation for mass segmentation.\\

        \cite{ribeiro2022deep}   & SE & Comparative study on different segmentation models. \\

        \cite{gudhe2022area}   & SE & Weight-adaptive multi-task learning network based on U-Net structure for pixel-wise breast density estimation.\\

        \cite{hwang2016self}   & CL, DE & CNN with a classification branch and a weakly-supervised localization branch.\\

        \cite{ben2017domain} & CL, DE & ROI extraction + CNN + clustering for classification; proposal generation + R-CNN for localization.\\

        \cite{al2017detection}  & DE, CL & YOLO for both detection and classification. \\

        \cite{ribli2018detecting}  & DE, CL & Faster-RCNN for breast lesion detection and classification.\\
        
        \cite{xiao2019learning} & DE, CL  & Faster-RCNN + top likelihood loss + similarity loss for cancer classification and mass detection.\\
        
        \cite{bakalo2019classification}   & CL, DE & Two-branch network with cancer classification and weakly-supervised detection.\\

        \cite{sainz2020multi}    & CL, DE & RetinaNet + Image-level classification for breast lesion detection and classification.\\

        \cite{boot2020diagnostic}   & DE, SE & Faster-RCNN + FPN~\cite{lin2017feature} + PANet~\cite{liu2018path} + LibraNet~\cite{pang2019libra} for mass detection; Recurrent Residual U-Net~\cite{alom2019recurrent} for mass segmentation.\\

        \cite{al2020deep}  & DE, SE, CL & YOLO for detection; full resolution convolutional network (RfCN) for segmentation; Inception-ResNet-V2 for classification.\\

        \cite{min2020fully}  & DE, SE & Pseudo-color processing to enhance mass-like pattern + Mask R-CNN for mass detection and segmentation.\\
       
        \cite{yi2021deepcat}  & CL, DE & ResNet34 for breast cancer classification; RetinaNet for mass detection.\\

        \cite{pinto2021impact} & CL, DE & AI system to improve radiologists in finding cancers from mammograms. \\ 

        \cite{li2021dual} & SE, CL & Multi-task CNN for simultaneous mass segmentation and breast cancer classification.\\

        \cite{baccouche2022early}  & CL, DE & YOLO-based fusion with GAN-based image synthesis for lesion detection and classification on prior images. \\

        \cite{jiang2022breast} & CL, DE & A three-stage deep learning detector for breast cancer classification and detection.\\

        \cite{jiang2019synthesize} & OT & Gradient guided conditional GAN-based (GGGAN) mammogram synthesis from DBT. \\

        \cite{jiang2021synthesis}   & OT & cGANs-based network for contrast-enhanced spectral mammography synthesis from low-energy mammograms. \\

        \cite{jiang2021TMIsynthesis}    & OT & GGGAn + perceptual loss for mammogram synthesis from DBT.\\

		\bottomrule
	\end{tabularx}
\end{table*}

\begin{table*}[t!]
	\centering
        \scriptsize
        \caption{Overview of deep learning-based ultrasound screening and diagnosis. CL=Classification, SE=Segmentation, DE=Detection.}
 	\label{tab:screening_and_diagnosis_utrasound}
	\begin{tabularx}{\textwidth}{p{0.06\textwidth}p{0.05\textwidth}p{0.75\textwidth}<{\raggedright}}
		\toprule
		\textbf{Reference} & \textbf{Task(s)} &  \textbf{Description} \\
		\midrule
\cite{cheng2016computer}                         & CL & Stacked denoising \textcolor{black}{auto-encoder} takes ROI as input for benign and malignant tumor classification.                                                                                                                                                                                                                                                              \\

\cite{antropova2017deep} & CL & Multi-level features from a pre-trained CNN + handcrafted features + SVM classifiers for benign and malignant classification. \\

\cite{han2017deep}                               & CL  & GoogLeNet \cite{szegedy2015going} to classify tumor ROIs into benign and malignant.                                                                                                                                                                                                                                                                                   \\
\cite{becker2018classification}                  & CL& Deep learning \textcolor{black}{software (ViDi Suite v. 2.0; ViDi Systems Inc, Villaz-Saint-Pierre, Switzerland)} is applied to classify benign and malignant tumors.                                                                                                                                                                                                                                                                            \\
\cite{hijab2019breast}                           & CL & Pre-trained VGG16 network with transfer learning for benign and malignant tumor classification.                                                                                         \\

\cite{al2019deep}                                & CL  & Combination with traditional data augmentation and GAN-based augmentation improves accuracy for classification of normal, benign, and malignant categories.                                         \\
\cite{moon2020computer}                          & CL & Model ensemble (VGGNet, ResNet, and DenseNet) for improving the classification performance.                                                                                            \\
\cite{cao2020breast}                             & CL  & Noise filter network for handling the \textcolor{black}{noisy biopsy labels}: two softmax layers for preventing deep models from overfitting noisy labels, and a teacher–student module for distilling the knowledge of clean labels.                                                                                                                    \\
\cite{zheng2020deep}                             & CL & ResNet50 pre-trained on ImageNet takes tumor ROIs as input for axillary lymph node status classification.                                                               \\

\cite{sun2020deep}                               & CL & DenseNet121 combined with molecular subtype information for classification of the axillary lymph node status.                                                            \\

\cite{zhou2020lymph}                             & CL & Inception-V3, Inception-ResNet-V2, and ResNet101 pre-trained on ImageNet for axillary lymph node status classification.   \\

\cite{xing2020using}                               & CL & \textcolor{black}{BI-RADS Vector-Attention Network (BVA Net) that trains with both texture information and decoded information from BI-RADS stratifications for breast masses classification.}   \\

\cite{guo2020deep}                               & CL & Two deep learning models pre-trained on ImageNet take tumor ROIs as input for prediction of risk of sentinel lymph node (SLN) and classification of Non-SLN metastasis, respectively.                                \\

\cite{erouglu2021convolutional}                               & CL & \textcolor{black}{Alexnet, MobilenetV2 and ResNet50 as base models + mRMR (Minimum Redundancy Maximum Relevance) feature selection + SVM to classify ultrasound images into normal, benign and malignant.}   \\

\cite{huang2021aw3m}                               & CL & \textcolor{black}{An auto-weighting and recovery framework for breast cancer classification based on four types of sonography.}   \\

\cite{lee2021axillary}                           & CL & Mask R-CNN for segmentation of the primary breast tumor and peritumoral regions + DenseNet121 for classification of axillary lymph node statuses.                                                                                                                                                                                               \\
\cite{zhu2021generic}                            & CL & Data augmentation to enrich ROI samples and VGG19 to classify benign and malignant tumors.\\

\cite{pang2021semi}                              & CL & Semi-supervised GAN for \textcolor{black}{data augmentation} + Inception-V3 for tumor ROI classification.  \\

\cite{wang2021deep}                              & CL & Combining two SSL methods, virtual adversarial training and FixMatch to boost the performance of recognition of benign and malignant tumors.                                                                                                                                                                     \\
\cite{qian2021prospective}                       & CL & Image pre-processing + B-mode, color Doppler, and elastography images as input + multimodal fusing model for breast cancer risk assessment + Grad-CAM for interpretation + Reader study and AI-assisted reader study.                                                                                       \\
\cite{chen2021domain}                            & CL & Contrast-enhanced ultrasound(CEUS) video and dual-amplitude CEUS image as input; data augmentation; 3D CNN backbone + domain-knowledge-guided temporal attention module + domain-knowledge-guided channel attention module.                                                                                            \\
\cite{sun2022deep}                               & CL & A custom CNN is built to classify tumor ROIs into different axillary lymph node statuses.      \\         

\cite{lu2022safnet}                               & CL & \textcolor{black}{Pre-trained ResNet-18 embedded with the spatial attention + majority voting to produce final classification results.}      \\

\cite{zhang2022semi}                             & CL  &  MixMatch for semi-supervised learning + model ensemble by voting.                                                                                                                                                                                                                \\
\cite{wang2022semi}                              & CL & A semi-supervised learning method for tumor classification: supervised learning + consistency training + adaptive token sampling.                                                                                                                                                                                           \\
\cite{kim2022weakly}                             & CL & Classification network + branch network learning segmentation information generated from reports for breast cancer classification.    \\

\cite{yap2017automated}                          & DE & Comparing a patch-based LeNet, U-Net, and a pre-trained FCN-AlexNet with transfer learning for tumor detection.                                                                                                                                                                                                                            \\
\cite{chiang2018tumor}                           & DE & 3D sliding window + 3D CNN for tumor candidate detection + hierarchical clustering for candidate aggregation + multi-scale tumor VOI aggregation to refine tumor detection results.                                                                                                                                                     \\
\cite{wang2019deeply}                            & DE & A deeply-supervised 3D U-Net-like network + threshold loss for cancer detection.                                                                                                                                                                     \\
\cite{yap2020breast}                             & DE & Faster-RCNN with Inception-ResNet-V2 + transfer learning + data augmentation for tumor detection.                                                                                                                                                                                                                                                                                        \\
\cite{lei2018segmentation}                       & SE & Deep boundary supervision with boundary cues + adaptive domain transfer for breast anatomical layer segmentation.                                                                                                                                                      \\
\cite{almajalid2018development}                  & SE & Breast tumor segmentation with U-Net + image contrast enhancement and speckle reduction + post-processing to removing noisy regions.                                                                                                                                                                              \\

\cite{zhuang2019rdau}                            & SE & U-Net + residual units + dilated convolutions + Attention Gate for tumor segmentation.                                                \\

\cite{xu2019medical}                             & SE & A custom 8-layer CNN for segmenting the breast ultrasound images into all major functional tissues automatically (skin, glands, and masses).                                                                                                                                                                                       \\
\cite{vakanski2020attention}                     & SE & U-Net + attention blocks + confidence level calculation for breast tumor segmentation.                                                                                                                               \\

\cite{xing2020lesion}                            & SE & Semi-pixel-wise cycle generative adversarial net with U-Net and ResNet for tumor segmentation.                                                                                                                                                                                              \\
\cite{zhu2020second}                             & SE & Attention-weighted sub-region pooling + guided multi-dimension second-order pooling for breast lesion segmentation.                                                                                                        \\

\cite{gomez2020comparative}                     & SE  & \textcolor{black}{Comparison of four public CNN-based semantic segmentation models (FCN with AlexNet network, U-Net, SegNet using VGG16 and VGG19 networks, and DeepLabV3+ using ResNet18, ResNet50, MobileNet-V2, and Xception networks) for breast tumor segmentation in ultrasound.}                                                                                                        \\

\cite{webb2021comparing}                             & SE & \textcolor{black}{An inter-observer reliability study comparing the performance of deep learning segmentation model against three experts.}                                                                                                        \\
\cite{wu2021bgm}                                 & SE & Feature pyramid network + boundary-guided feature enhancement + multi-scale image information for tumor segmentation.    \\     
\multicolumn{3}{r}{\footnotesize(\textit{continued on the next page})}\\

\end{tabularx}
\end{table*}

\begin{table*}[t!]
\ContinuedFloat
	\centering
        \scriptsize
        \caption{Overview of deep learning-based ultrasound screening and diagnosis. CL=Classification, SE=Segmentation, DE=Detection.}
	\begin{tabularx}{\textwidth}{p{0.06\textwidth}p{0.05\textwidth}p{0.75\textwidth}<{\raggedright}}
		\toprule
		\textbf{Reference} & \textbf{Task(s)} &  \textbf{Description} \\
		\midrule

\cite{zhou2021cross}                                 & SE                                                  & \textcolor{black}{Cross-model attention-guided V-Net with an improved Mask R-CNN head for 3D automated breast
ultrasound segmentation.}        \\

\cite{zou2021robust}                             & SE & Noisy annotation tolerance network with noise index + noise index combined with a dynamic index loss function to suppress the influence of noise for tumor segmentation.                                                                                              \\
\cite{ning2021smu}                               & SE  & Saliency-guided morphology-aware U-Net with saliency maps + background-assisted fusion, shape-aware, edge-aware, and position-aware units for lesion segmentation.\\

\cite{tong2021improved}                          & SE & Improved U-Net based on Mixed Attention Loss Function model for breast tumor segmentation.                                                                                          \\
\cite{wang2021residual}                          & SE  & ResNet34-based encoder-decoder structure + residual representation module + residual feedback transmission for breast lesion segmentation.                                                                                                                                                                                 \\
\cite{xue2021global}                             & SE  & CNN with atrous spatial pyramid pooling + global guidance block and lesion boundary detection module for breast lesion segmentation.                                                                          \\

\cite{ferreira2022comparative}                   & SE  & Comparative study on tumor segmentation, including U-Net, Dynamic U-Net, Semantic Segmentation Deep Residual Network with Variational Autoencoder, U-Net Transformers, Residual Feedback Network, Multi-scale Dual Attention-Based Network, and Global Guidance Network.                                                         \\
\cite{chen2022c}                                 & SE  & A cascaded convolutional neural network integrating U-Net, bidirectional attention guidance network, and refinement residual network for breast tumor segmentation.                                                                \\

\cite{meng2022ceusegnet}                         & SE & A U-Net-like network with dual top-down branches and residual connections for breast lesion segmentation from contrast-enhanced ultrasound.                                                                                                 \\
\cite{zhai2022ass}                               & SE & Asymmetric semi-supervised GAN with two generators and a discriminator for tumor segmentation.                                                                                                             \\
\cite{li2022deepUS}                                & SE & Two-step pipeline: 1) semi-supervised semantic segmentation for breast anatomy decomposition; 2) tumors are recognized by a classification model and then segmented by class activation mapping and deep level set method. \\

\cite{kim2021weakly}                             & CL,  DE  & VGG16, ResNet34, and GoogLeNet for classification + class activation map for weakly-supervised tumor localization. 

\\

\cite{li2021weakly}                              & CL, SE  & ResNet101 or Xception with atrous spatial pyramid pooling \cite{chen2018encoder} + class activation map with level set for weakly-supervised tumor segmentation.                                                                                                                                                          \\
\cite{ragab2022ensemble}                         & SE, CL  & Image pre-processing by wiener filtering and contrast enhancement + Chaotic Krill Herd Algorithm with Kapur’s entropy for tumor segmentation; An ensemble of VGG16, VGG19 and SqueezeNet for feature extraction + Cat Swarm Optimization with multilayer perceptron for cancer classification. \\

\cite{shin2018joint}                             & DE, CL & Joint weakly- and semi-supervised method with Faster R-CNN for localization and classification of masses.                                                                      \\

\cite{cao2019experimental}                       & DE, CL  & Comparative study on using state-of-the-art methods for breast lesion detection and classification.\\

\cite{tanaka2019computer}                        & CL, DE  & Ensemble network (VGG19 and ResNet152) for single view + fusing of view-level results for mass-level classification and detection.\\

\cite{liao2019automatic}                               & SE, CL   & \textcolor{black}{Tumor regions segmented using the supervised block-based region segmentation algorithm + a VGG19 network pretrained on the ImageNet to predict benign or malignant lesions.}  \\

\cite{misra2022deep}                               & CL,SE & \textcolor{black}{A weighted multimodal U-Net model for segmenting lesions + a multimodal fusion framework on cropped B-mode and strain elastography (SE)-mode lesion images to classify benign and malignant lesions.}      \\ 

\cite{huang2022boundary}                         & SE, CL & A boundary-rendering framework performs segmentation and classification simultaneously.                                 \\

\cite{vigil2022dual}                             & SE, CL & Convolutional deep autoencoder model to segment and extract deep radiomics of the breast lesions + a Random Forest for benign and malignant tumor classification. \\

\cite{podda2022fully}                            & SE, CL & Specialized CNN ensembles + cyclic mutual optimization for simultaneous classification and segmentation. \\

\cite{zhou2021multi}                             & SE, CL  & Modified 3D U-net for tumor segmentation + lightweight multi-scale network for classification + iterative feature refinement.                                                 \\
  	\bottomrule
	\end{tabularx}
\end{table*}

\begin{table*}[!t]
	\centering
        \scriptsize
        \caption{Overview of deep learning-based MRI diagnosis. CL=Classification, SE=Segmentation, DE=Detection, OT=Other Tasks; ID=In-house Dataset; MIP=Maximum Intensity Projection; DCE=Dynamic Contrast-Enhanced MRI; T1=T1-weighted MRI; T1+C=Contrast-enhanced T1 MRI; NFS=non-fat saturated; T2=T2-weighted MRI; DWI=Diffusion-weighted Imaging; ADC=Apparent Diffusion Coefficient.}
 	\label{tab:screening_and_diagnosis_mri}
	\begin{tabularx}{\textwidth}{p{0.06\textwidth}p{0.04\textwidth}p{0.12\textwidth}p{0.62\textwidth}<{\raggedright}}
		\toprule
		\textbf{Reference} & \textbf{Task(s)} & \textbf{Sequence} &  \textbf{Description} \\
		\midrule 

        \cite{rasti2017breast} & CL & DCE & a mixture ensemble of convolutional neural networks for benign and malignant classification.   \\
        
        \cite{antropova2017deep} & CL & DCE & Multi-level features from a pre-trained CNN + handcrafted features + SVM classifiers for benign and malignant classification. \\
        
        \cite{antropova2018use} & CL & DCE &  Comparison of the maximum intensity projection (MIP)-based and other center slides-based classification with pre-trained VGGNet + SVM classifier.  \\
        
        \cite{zheng2018small} & CL & DCE & Dense convolutional LSTM for benign and malignant classification.\\
                
        \cite{whitney2019comparison} & CL & DCE & Compare CNN-based transfer learning, hand-engineered feature-based radiomics, and fusion classifiers.   \\ 
        
        \cite{herent2019detection} & CL & T1  & Supervised-attention model with deep learning for breast lesion detection and classification.\\
        
        \cite{fujioka2021deep} & CL & DCE & Comparison of different CNNs on benign and malignant classification with MIP.\\

        \cite{zhou2020diagnosis} & CL & DCE  & Comparison of radiomics model and ResNet50 for malignant and benign classification. \\

        \cite{feng2020knowledge} & CL & DCE, DWI  & Knowledge-driven Feature Learning and Integration (KFLI) framework for benign and malignant classification. \\

        \cite{hu2020deep} & CL &  DCE,T2 & Pre-trained CNN + SVM classifier + multimodal fusion for benign and malignant classification. \\
                
        \cite{zhang2021prediction} & CL & DCE & CNN + convolutional LSTM + transfer learning for molecular subtyping.\\
        
        \cite{liu2021preoperative} & CL & T1+C,T2, DWI & CNN + transfer learning + multilayer perceptron classifier for Ki-67 status prediction. \\ 

        \cite{chen2022deep} & CL & DCE & Faster R-CNN for ROI localization + CNNs for benign and malignant classification. \\

        \cite{sun2021prediction} & CL & DCE& Ensemble CNNs for molecular subtyping. \\

        \cite{hu2021improved} & CL & DCE & CNN feature extractor + SVM classifier + MIP for benign and malignant classification. \\

        \cite{verburg2022deep} & CL & DCE & CNN-based model for screening and diagnosis of extremely dense breasts.\\ 
        
        \cite{witowski2022improving} & CL & DCE  & 3D ResNet18 with max pooling layer before linear classifier + pre-trained weights from video dataset for malignancy classification. \\ 

        \cite{jing2022using} & CL & Ultrafast & Modified ResNet34 with pre-trained weights for breast cancer screening from ultrafast MRI.  \\ 

        \cite{ren2022convolutional} & CL & T1,T2,DCE  & Comparing five CNN models based on different MRI modalities for detecting axillary lymph node metastasis. \\

        \cite{yin2022combined} & CL & DCE, ADC, T2 & Pre-trained ResNet18 for differentiating triple-negative breast cancer from fibroadenoma BI-RADS 4 lesions. \\ 

        \cite{liu2022weakly} & CL & T1+c  & ResNet101 for benign and malignant classification.\\ 

        \cite{sun2022transfer} & CL & DCE & Compare unsupervised pre-training and ensemble learning for transfer learning-based molecular subtyping. \\

        \cite{jing2022breast} & CL & DCE, Ultrafast DCE & Modified ResNet18 with pre-trained weights + LSTM model for benign and malignant classification.   \\

        \cite{zhang2018hierarchical} & SE, CL & DCE & Mask-guided hierarchical learning based on fully-convolutional networks for coarse-to-fine tumor segmentation + identification of luminal A subtype from segmentation masks.\\

        \cite{piantadosi2019dce} & SE & DCE & U-shaped models that exploits the  Three Time Points (3TP) approach for lesion segmentation. \\
        
        \cite{galli2021pipelined} & SE & DCE & Breast masking + 3TP slice extraction + motion correction + modified U-Net for lesion segmentation. \\
        
        \cite{yue2022deep}& SE & DCE & Residual U-Net for cancer segmentation and measurement. \\

        \cite{vidal2022u} & SE & DCE & U-net ensemble for breast lesion segmentation.  \\

        \cite{maicas2017deep} & DE & DCE & Deep Q network with attention mechanism for breast lesion detection. \\ 
        
        \cite{dalmics2018fully} & DE & DCE & Lesion candidate detection with U-Net + contralateral input patch + CNN for candidate classification. \\
        
        \cite{ayatollahi2021automatic}  & DE & Ultrafast DCE & Modified 3D RetinaNet for breast lesion detection.\\

        \cite{zhou2019weakly} & DE, CL & DCE &  Deep 3D DenseNet for benign and malignant classification + CAM and Dense Conditional Random Field for weakly-supervised localization.\\

        \cite{lu2019breast} & SE, CL & T1, T2, DWI, DCE & Four-mode linkage backbone with refined U-Net on different modalities for segmentation; CNN for classification. \\

        \cite{luo2019deep} & DE, CL & DCE & Cosine Margin Sigmoid Loss for benign and malignant classification + Correlation Attention Map for weakly-supervised tumor detection.   \\

        \cite{parekh2020multiparametric} & SE, CL & T1,T2,DWI,DCE & Stacked sparse autoencoder networks for segmentation combined with SAE-SVM for classification. \\

        \cite{jiao2020deep} & SE, DE & DCE &  U-Net++ for breast segmentation; Faster-RCNN for mass detection on segmented images.  \\

        \cite{zhu2022development} & SE, CL & DCE, DWI & V-Net for lesion segmentation from DCE, Attention U-net for lesion segmentation from DWI, ResNet-based model with both DCE and DWI inputs for benign and malignant classification.  \\
        
        \cite{chung2022deep} & OT & T1,T2, DWI,ADC & 3D CNN-based model for generating simulated multiparametric MRI. \\
        
        \cite{van2020volumetric} & OT & DCE & 3D regression CNN for breast density estimation. \\ 

		\bottomrule
	\end{tabularx}
\end{table*}

\begin{table*}[]
  \centering
  \scriptsize
  \caption{Overview of deep learning-based pathology diagnosis. CL=Classification, SE=Segmentation, DE=Detection, SN=Stain Normalization, VS: Virtual Staining.}
  \label{tab:screening_and_diagnosis_pathology}%
  \begin{tabularx}{\textwidth}{p{0.06\textwidth}p{0.05\textwidth}p{0.8\textwidth}<{\raggedright}}
  \toprule
  \textbf{Reference} & \textbf{Task(s)} &  \textbf{Description} \\
  \midrule
  
  \cite{hatipoglu2014classification}    & CL & Customized CNN for breast cancer classification.\\

  \cite{wang2016deep}  & CL    & Otsu algorithm for pre-processing + patch-based classification + heatmap-based post-processing for whole-slide image classification. \\
  
  \cite{bejnordi2017diagnostic}   & CL  & Assessing different deep learning methods on detection of lymph node metastases and comparing the methods with pathologists.\\

  \cite{vandenberghe2017relevance}    & CL & Cell detection using colour deconvolution + cell classification using CNN.\\
  
 \cite{araujo2017classification}    & CL  &  Data augmentation + CNNs for patch-wise classification. \\
  
  \cite{wan2017automated}    & CL & Stain normalization + active contour (AC)-based segmentation method for nuclei segmentation + CNN to extract semantic features + pixel-level (texture) and object-level (architecture) features + multiple SVMs for breast cancer classification.    \\
    
  \cite{bejnordi2017context}    & CL  & A CNN for classification of high-resolution patches + context-aware stacked CNNs for producing dense prediction maps + a WSI labeling module.   \\ 
  
  \cite{ilse2018attention}    & CL & Attention-based deep multiple instance learning for H\&E stained WSI classification.  \\
        
  \cite{ehteshami2018using}    & CL  & Stack CNNs: Network I to identify stromal regions + Network II to identify cancer-associated stroma + Network III to predict WSI-level probability of invasive cancer. \\
  
  \cite{lee2018robust}    & CL &  Patch-level CNN for metastasis detection + slide-level lymph node classifier for pN-stage prediction. \\
  
  \cite{qi2018label}    & CL &  Active learning with an entropy-based strategy and a confidence-boosting strategy for breast cancer classification. \\
  
  \cite{peikari2018cluster}    & CL & A cluster-then-label method to identify high-density regions + SVM for semi-supervised breast cancer classification. \\
  
  \cite{couture2018image}    & CL & Color and intensity normalization + VGG16 pre-trained on ImageNet + SVM classifier ensemble for breast cancer classification.\\

  \cite{feng2018breast}    & CL & Stacked denoising autoencoder for cancer cell nuclei classification.    \\
  
  \cite{campanella2019clinical}    & CL & Multiple instance learning (MIL) for feature extraction + recurrent neural network (RNN)-based information integration for final classification. \\
  
  \cite{khan2019novel}    & CL & GoogLeNet, VGG, and ResNet with transfer learning for feature extraction + fully connected layer for final classification.\\
    
  \cite{sudharshan2019multiple}    & CL & Comparing different MIL-based algorithms, such as KNN-based, SVM-based, and deep learning-based approaches, on breast cancer classification. \\

  \cite{turkki2019breast} & CL & Use VGG16 to classify patients into low/high digital risk score groups\\

  \cite{mercan2019assessment}    & CL & Developing tissue distribution feature and structure feature from CNN-segmentation results for differentiating biopsy types (benign, atypia, ductal carcinoma in situ, and invasive cancer.). \\
       
  \cite{su2019local}    & CL & Local and global consistency regularized Mean Teacher for semi-supervised nuclei classification. \\
  
  \cite{alom2019breast}    & CL & Inception recurrent residual convolutional neural network for breast cancer classification. \\
  
  \cite{naik2020deep}    & CL & Attention-based MIL + data augmentation for determining estrogen receptor status from H\&E-stained WSIs.   \\
  
  \cite{he2020integrating}    & CL & ST-Net (ImageNet pre-trained DenseNet-121) for the prediction of local gene expression from H\&E-stained histopathology images. \\
  
  \cite{valieris2020deep}    & CL & CNN-based patch-wise classification + RNN-based information aggregation for DNA repair deficiency prediction. \\

  \cite{noorbakhsh2020deep}    & CL & Inception-V3 for tumor/normal, cancer subtype, and mutation classification. \\
  
  \cite{sheikh2020histopathological}    & CL & Multi-scale input and multi-feature network for breast cancer classification. \\
  
  \cite{teh2020learning}    & CL &  Learning transferable features from multiple organs with proxy-based metric learning approach from weakly labeled data for breast cancer classification. \\

  \cite{li2021dualwsi}    & CL & Self-supervised learning for feature extraction + pyramidal fusion mechanism for multi-scale WSI feature + MIL aggregator with attention for WSI cancer classification.    \\

  \cite{yu2021deep}    & CL & Proposing a DL-empowered breast cancer auxiliary diagnosis scheme for remote e-health supported by 5G technology. \\

  \cite{faryna2021tailoring}   & CL & Domain-specific data augmentation + CNN for breast cancer classification. \\
  
  \cite{wetstein2021deep}    & CL & DenseNet-121-based automated ductal carcinoma in situ grading in breast histopathology images. \\
  
  \cite{li2021deep}    & CL  & Multiple CNNs for tumor epithelium identification and pathological complete response score prediction.    \\
  
  \cite{mi2021deep}    & CL & CNN-based two-stage architecture for multi-class classification (normal tissue, benign lesion, ductal carcinoma in situ, and invasive carcinoma). \\
  
  \cite{wang2021prediction}    & CL & ResNet18 for predicting gBRCA mutation risk.    \\
  
  \cite{senousy2021mcua}    & CL & Multi-level context-aware models + uncertainty aware dynamic model ensemble for breast cancer classification. \\
  
  \cite{wang2022improved}    & CL & An ensemble consisting of 20 Inception V3 models for Nottingham histological grading.    \\
  
  \cite{mantrala2022concordance}    & CL & Compare AI with a group of breast pathologists on breast carcinoma grading.\\
  
  \cite{mercan2022deep}    & CL & Deep regression model for nuclear pleomorphism scoring (the extent of abnormalities in the overall appearance of tumor nuclei).   \\
  
  \cite{jaroensri2022deep}    & CL & Two-stage system: CNNs for patch-level classification + lightweight slide-level classifier for breast cancer histological scoring.    \\
  
  \cite{wetstein2022deep}    & CL & MIL with ResNet34 + multi-task learning for tumor grading.\\

  \cite{saini2022vggin}    & CL    & Transfer learned knowledge of VGG16 from ImageNet to the smaller imbalanced breast cancer dataset.   \\

  \cite{zhang2022dtfd}    & CL  & Attention for instance probability derivation + pseudo bag-based slide-level breast cancer analysis.   \\

  \cite{schirris2022deepsmile}    & CL  & Self-supervised contrastive learning for feature extractor pre-training + variability-aware MIL for WSI-based classification.\\
  
  \cite{mehta2022end}    & CL & Transformer-based holistic attention network to encode global information for breast biopsy images diagnosis.\\
  
  \cite{lu2022slidegraph}    & CL  & Graph neural network (GNN)-based model (SlideGraph+) to predict HER2 status.\\

  
  \cite{cirecsan2013mitosis}    & DE   & Max-pooling CNN + ensemble for mitosis detection. \\

  \cite{irshad2014spectral}    & DE & Focal plane selection + spectral bands selection + candidate detection and feature computation + MLP classifier for mitosis detection. \\
  
  \cite{romo2016automated}    & DE  & Blue ratio transform for nuclei candidates detection + patch-level DNN classifier for tubule formation quantification.    \\

  \cite{kong2017cancer}    & DE  & Spatially Structured Network that combines CNN and 2D LSTM to detect the metastasis locations. \\
  
  \cite{chen2016mitosis} & DE   & Cascaded CNNs: a coarse retrieval model for mitosis candidate detection + a fine discrimination model for singling out final mitosis results.\\
  
  \cite{li2018deepmitosis} & DE & Multi-stage deep learning framework with detection, verification, and segmentation. \\
  
  \cite{li2019weakly} & DE  & Concentric loss for weakly supervised mitosis detection. \\
  
  \cite{balkenhol2019deep} & DE  &  CNN to assist observers in mitotic detection from breast cancer WSI. \\

  \multicolumn{3}{r}{\footnotesize(\textit{continued on the next page})}\\
  \end{tabularx}%
  \end{table*}

  \begin{table*}[]
  \ContinuedFloat
  \centering
  \scriptsize
  \caption{Overview of deep learning-based pathology diagnosis. CL=Classification, SE=Segmentation, DE=Detection, SN=Stain Normalization, VS: Virtual Staining.}
  \begin{tabularx}{\textwidth}{p{0.05\textwidth}p{0.08\textwidth}p{0.8\textwidth}<{\raggedright}}
  \toprule
  \textbf{Reference} & \textbf{Task(s)} &  \textbf{Description} \\
  \midrule
  \cite{jin2020integrative} & DE &  Detection of lymph node metastases via U-net with additional segmentation channels of the nucleus, mitotic figures, epithelium, and tubules. \\
      
  \cite{sebai2020maskmitosis} & DE & Mask RCNN for mitosis detection. \\

  \cite{mahmood2020artificial} & DE & Multi-stage detection based on Faster-RCNN and CNNs for mitosis detection. \\
  
  \cite{lei2020attention} & DE & Attention-guided multi-branch CNN for mitosis detection. \\
   
  \cite{sohail2021mitotic} & DE & Mask R-CNN for candidate mitoses selection + ensemble-based mitoses classification for mitotic nuclei analysis. \\
  
  \cite{wang2022label}    & DE & Label Cleaning MIL for coarse annotations refinement based on a single WSI. \\
  
  \cite{wang2022generalizable}    & DE & Fourier-based data augmentation + Unet backbone + channel-wise multi-scale attention mechanism for mitosis detection. \\
  
  \cite{pan2017accurate}    & SE & Sparse reconstruction for accentuating the nuclei region + CNN cascaded by multi-layer CNNs for nuclei segmentation + morphological operations and prior knowledge for post-processing. \\
  
  \cite{kumar2017dataset}  & SE    & Color normalization + three-class CNN for patch-wise prediction + post-processing technique for Nuclear Segmentation.\\
  
  \cite{naylor2017nuclei}    & SE  & Data augmentation +  ensemble of segmentation networks + post-processing strategy for  cell nuclei segmentation.   \\
  
  \cite{naylor2018segmentation}   & SE  & Deep regression of the distance map with fully convolutional networks for nuclei segmentation. \\
     
  \cite{mehta2018learning} & SE & Encoder-decoder model with modifications: an input-aware encoding block + a new dense connection pattern + dense and sparse decoders + a multi-resolution network for breast tissue segmentation. \\

  \cite{liu2018end}    & SE & One FCN for all nuclei extraction + another FCN for tumor nuclei extraction + a multi-column CNN for H-score prediction.\\

  \cite{mahmood2019deep}    & SE & Adversarial training for nuclei segmentation from breast histopathology images. \\
  
  \cite{guo2019fast}    & SE & Cascade classification network and segmentation network for cancer region segmentation. \\
  
  \cite{chan2019histosegnet} & SE & Four-stage algorithm: 1) patch-level CNN for histological tissue type (HTT) classification; 2) pixel-level HTT segmentation; 3) Inter-HTT adjustment; and 4) HTT segmentation post-processing. \\

  \cite{yoo2019pseudoedgenet}    & SE &   Segmentation network + edge network + attention module for nuclei segmentation with point annotation.\\

  \cite{zhou2019cia}   & SE    & 
  Contour-aware Informative Aggregation Network (CIA-Net) with multilevel information aggregation module between two task-specific decoders for nuclei segmentation. \\

  \cite{xie2020instance}    & SE & Self-supervised learning by leveraging the prior knowledge of nuclei size and quantity for nuclei segmentation.  \\
  
  \cite{lee2020scribble2label}    & SE &  Weakly supervised cell segmentation by generating reliable pseudo labels from scribbles. \\
  
  \cite{sahasrabudhe2020self}    & SE & Self-supervised signal + fully convolutional attention network + regularization constraints for nuclei segmentation. \\
  \cite{tian2020weakly}    & SE &  Point label + coarse segmentation mask generation based on clustering + geometric-constrained negative boundaries for weakly supervised nucleus segmentation.\\
  
  \cite{xiang2020bio}    & SE & A recurrent bi-directional O-shape network for nuclei segmentation. \\
  
  \cite{qu2020weakly}   & SE & Semi-supervised nuclei detection with partial points + Voronoi diagram and cluster label for coarse label generation + weakly-supervised nuclei segmentation + dense CRF for post-processing. \\
 
  \cite{wang2020improved}   & SE  & An improved U-Net++ architecture for nuclei segmentation. \\
  
  \cite{wan2020robust}   & SE & A modified atrous spatial pyramid pooling U-Net + a weighted binary cross entropy loss function with Dice loss + boundary refinement for nuclei segmentation. \\
    
  \cite{priego2020automatic}    & SE & CNN-based processing pipeline for segmentation from breast histopathology images.\\
  
  \cite{ho2021deep}    & SE &  Multi-magnification network for segmenting multiple tissue subtypes on breast pathology images. \\
    
  \cite{zhao2021net}   & SE & Multi-task network with attention for nuclei segmentation. \\
  
  \cite{he2021cdnet}    & SE &  A Centripetal Direction Network for nuclear instance segmentation.  \\
  
  \cite{valanarasu2021medical} & SE & Gated axial-attention model with local and global branches for nuclei segmentation. \\
  
  \cite{liu2021mdc}    & SE & A multi-scale connected network with distance map and contour information for nucleus segmentation\\
  
  \cite{zeiser2021deepbatch}    & SE & Pre-processing + U-Net for ROI detection + ROI sampling + ResNet50/U-Net for breast cancer region segmentation. \\

  \cite{wang2022uctransnet} & SE &  Modified U-Net with a channel transformer for medical image segmentation. \\
  
  \cite{jin2022semi}    & SE &  A Mean Teacher-based hierarchical consistency enforcement framework + a hierarchical consistency loss for semi-supervised histological image segmentation.  \\

  \cite{wang2022smeswin}    & SE & Merging CNN and Transformer for medical image segmentation. \\

  \cite{li2022rt}    & SE & Combining Unet, Transformer, and residual network for nuclei segmentation. \\
  
  \cite{lin2022insmix}   & SE  & Data augmentation with adversarial learning + instance segmentation network for nuclei segmentation. \\

  \cite{lin2023nuclei}   & SE & Co-trainng and self-supervised learning for weakly-supervised nuclei segmentation. \\
  
  \cite{bentaieb2017adversarial}    & SN, CL, SE  & 
  Stain transfer network with adversarial training +task-specific network for stain normalization, and image classification, and segmentation.\\

  \cite{tellez2018h}    & SN, DE & A data augmentation strategy in the H\&E color space to simulate H\&E stain variations for improving mitosis detection. \\ 
  
  \cite{tellez2018whole}    & SN  &  Data augmentation + knowledge distillation to train stain-invariant network for mitosis detection.\\
    
  \cite{alirezazadeh2018representation}    & SN, CL  & A representation learning-based unsupervised domain adaptation for classification of breast cancer.\\
  
  \cite{anghel2019high}    & SN,CL   & Parameter cross-checking technique to enhance stain normalization pipeline for stain normalization. \\

  \cite{shrivastava2021self}    & SN  & Self-attentive adversarial stain normalization for mapping multiple stain appearances to a common domain.     \\
    
  \cite{cruz2017accurate}    & CL, DE &  Tile tissue sampling + tile pre-processing + CNN for breast cancer classification. \\

  \cite{saha2018her2net}    & CL, DE, SE & Integrating U-Net and trapezoidal long short-term memory (TLSTM) for cell membrane and nucleus detection, segmentation, classification and HER2 scoring. \\
  
  \cite{mehta2018net}    & CL, SE & Modified U-net with discriminative map generation and convolutional block modularity for breast cancer classification and segmentation. \\

  \cite{graham2019hover}   & SE, CL & A new CNN architecture for simultaneous nuclear segmentation and classification. \\
  
  \cite{mahanta2021ihc}    & SE, CL &   A fully convolutional neural network for automated nuclear segmentation and ensemble classification for Allred scoring. \\

  \cite{breen2021assessing}    & SN, DE  &  Evaluation of Cycle GAN and Neural Style Transfer for domain adaptation in mitosis detection. \\  

  \cite{ghahremani2022deep}    & VS, SE, CL &  Multi-task network for stain deconvolution/separation, cell segmentation, and quantitative single-cell IHC scoring. \\

  \cite{graham2022one} & SE, CL & Multi-task learning for simultaneous histology image segmentation and classification. \\
  
  \cite{lafarge2017domain}    & SN, CL  &  Domain-adversarial neural network for mitosis detection. \\

  \cite{rivenson2018deep}    & VS &  Generative adversarial network for transforming an auto-fluorescence image of an unlabeled tissue section into an image that is equivalent to the bright-field image of the stained version of the same sample. \\
  
  \cite{bai2022label}    & VS &  A DL-based virtual HER2 IHC staining method using a conditional GAN. \\
  
  \cite{mercan2020virtual}    & VS, CL   & Conditional GAN and Cycle GAN for translation between PHH3 and H\&E images. \\
  
  \cite{liu2020global}    & VS, CL  & U-Net with global pixel transformer layers and dense blocks for virtual staining. \\
  
 \bottomrule
 \end{tabularx}%
 \end{table*}

\begin{table*}[!t]
    \centering
          \scriptsize
          \caption{Overview of deep learning-based treatment response prediction. CL=Classification, OT=other tasks NAC=Neoadjuvant Chemotherapy, pCR=pathological Complete Response, US=Ultrasound, MRI=Magnetic Resonance Imaging, DCE=Dynamic Contrast-Enhanced MRI, CT=Computed Tomography, MG=Mammograms, WSI=Whole Slide Image, H\&E=Hematoxylin and Eosin stained pathology image, IHC=Immunohistochemistry pathology images. NIM=Non-Image Modalities}
     \label{tab:treatment response}
    \begin{tabularx}{\textwidth}{p{0.05\textwidth}p{0.02\textwidth}p{0.1\textwidth}p{0.7\textwidth}<{\raggedright}}
      \toprule
      \textbf{Reference} & \textbf{Task} & \textbf{Modality} &  \textbf{Description} \\
      \midrule 
             
          \cite{huynh2017comparison} & CL & DCE  & Multiple input of DCE images at different contrast time points with pre-trained VGG for predicting response to neoadjuvant chemotherapy. \\
  
          \cite{machireddy2018analysis} & CL & MRI  & CNN-based classification for NAC pCR classification based on pre-treatment DCE-MRI. \\
  
          \cite{ravichandran2018deep} & CL  & MRI & Multiple inputs of pre- and post-contrast MRI based on CNN models for prediction of neoadjuvant treatment response. \\ 
  
          \cite{ha2019prior} & CL & MRI  & Custom CNN-based frameworks for NAC pCR classification.\\

          \cite{choi2020early} & CL & CT/PET+ MRI &  CNN-based networks for NAC pCR classification using PET/CT and MRI.\\
          
          \cite{el2020multi} & CL & MRI & Multi-input parallel CNN combining features from the pre- and the post-treatment DCE-MR images for NAC pCR classification.\\
  
          \cite{liu2020novel} & CL & MRI   & CNN-based frameworks developed on multi-institutional data for NAC pCR classification. \\
  
          \cite{qu2020prediction} & CL & MRI  & CNN-based frameworks with inputs of multi-phase DCE-MRI of pre-NAC and post-NAC + molecular subtype information for NAC pCR classification.  \\

          \cite{duanmu2020prediction} & CL & MRI,NIM  & Convolutional neural network of post-contrast MRI + non-imaging clinical data (molecular and demographic data) for NAC pCR classification. \\

          \cite{braman2020deep} & CL & MRI & CNN-based frameworks from pre- and post-contrast MRI for HER2-targeted NAC for NAC pCR classification.  \\
          
          \cite{byra2020early} & CL & US  & Siamese CNNs model with pre-NAC and post-NAC image for NAC pCR classification.   \\
    
          \cite{comes2021early} & CL & MRI   & Pre-treatment and early-treatment DCE-MRI + pre-trained CNN for low-level feature extraction + SVM classifier for NAC pCR classification.\\
          
          \cite{li2021deep}   & CL & H\&E  & Multiple CNNs for tumor epithelium identification and pathological complete response score prediction.    \\
  
          \cite{joo2021multimodal} & CL & MRI, NIM  & Multimodal network with contrast-enhanced T1 MRI and T2 MRI + clinical information inputs for NAC pCR classification. \\
  
          \cite{jiang2021ultrasound} & CL & US, NIM & CNN for deep features extraction with pre- and post-treatment images + hand crafted features + clinical information for NAC pCR classification. \\
  
          \cite{duanmu2021spatial} & CL & H\&E + IHC & CNN-based tumor cell detection modules + spatial attention modules for NAC pCR classification based on multi-stained inputs.\\
  
          \cite{gan2021clinical} & CL & MRI & 3D-UNet for breast tumor and axillary areas segmentation + radiomics extraction +clinicopathologic factor + logistic regression for NAC pCR prediction.   \\
  
          \cite{peng2022pretreatment} & CL & MRI, NIM & CNN-based deep feature + kinetic parameters + molecular information for NAC pCR classification. \\
         
          \cite{farahmand2022deep} & CL  & H\&E WSIs & Pre-trained Inception-V3 for HER2 classification and trastuzumab response from H\&E images. \\

          \cite{taleghamar2022deep} & CL & US &  Residual network and residual attention network for feature extraction + fully connected network for NAC pCR classification using pre-treatment quantitative ultrasound (QUS) multiparametric imaging. \\

          \cite{byra2022prediction}  &CL & US  & CNN for feature extraction + recurrent network for NAC pCR classification from serial US images. \\
  
          \cite{qi2022multi}  & CL & CT & MultiResUnet3D \cite{ibtehaz2020multiresunet} + multi-layer perceptron for NAC pCR classification.  \\

          \cite{skarping2022analysis}  & CL & MG & Siamese CNN with tumor patch and reference image patch as inputs for NAC pCR classification.  \\
  
          \cite{li2022deep}  & CL & H\&E WSIs, NIM  & Pre-trained ResNet18 for patch feature extraction + logistic regression + clinical information for NAC pCR classification. \\

          \cite{wu2022integrated} & CL & US  & U-Net for tumor segmentation + signatures extraction + clinical factors + a deep learning-based serial ultrasonography assessment system for NAC pCR prediction. \\
          
          \cite{xie2022dual} & CL & US & Dual-branch CNN with pre- and post-NAC images as inputs for NAC pCR classification.\\
         
          \cite{gu2022deep} & CL & US & A deep learning radiomics pipeline for prediction of NAC pCR at different time points of treatments.  \\
  
          \cite{tong2022dual} & CL  & US & Dual-input transformer with pre- and post-NAC images for NAC pCR classification. \\

          \cite{liu2022early} & CL  & US & Siamese multi-task CNN for tumor segmentation and NAC pCR classification.\\
  
          \cite{aghaei2015computer} & OT & MRI  & Breast area and tumor segmentation + kinetic feature extraction + artificial neural network for classification and identifying clinical markers for NAC treatment response prediction.  \\

          \cite{li2022predicting} & OT & H\&E & Propose a CNN-based tumor-associated stroma score (TS-score) as new predictor for predicting pathological complete response (pCR) to NAC.\\

      \bottomrule
    \end{tabularx}
  \end{table*}
  
\begin{table*}[t]
	\centering
        \scriptsize
        \caption{Overview of deep learning-based prognosis. H\&E=Hematoxylin and Eosin stained pathology image; SR=Sirius Red stained pathology image; NIM = Non-Image Modalities; US=ultrasound; MG=mammogram; CL=Classification, SP=Survival Prediction, ReP=Recurrence Prediction, OT=other Tasks.}
 	\label{tab:prognosis}
	\begin{tabularx}{\textwidth}{p{0.05\textwidth}p{0.05\textwidth}p{0.17\textwidth}p{0.7\textwidth}<{\raggedright}}
		\toprule
		\textbf{References} & \textbf{Task(s)} & \textbf{Modality} &  \textbf{Description} \\
		\midrule
        
        \cite{liu2021tsdlpp} & SP & H\&E & Two-stage deep learning paradigm + cox regression model for survival time prediction.\\
        
        \cite{morkunas2021tumor} & SP & SR & ANN+cox regression model for survival time prediction.\\
        
        \cite{wetstein2022deep}  & SP & Pathology & MIL with ResNet34 + multi-task learning for tumor grading and survival analysis.\\
        
        \cite{wang2021gpdbn} &  SP &  Pathology images, NIM & deep bilinear network integrating both genomic data and pathological images for 5-year survival prediction.\\
        
        \cite{liu2022deep} & SP & H\&E + NIM & Divide survival time into intervals and classify the survival time via MobileNetV2.\\
        
        \cite{ha2019convolutional} & ReP & MRI & CNN to predict recurrence chance via classifying Oncotype Dx Recurrence Score groups. \\
        
        \cite{yu2020predicting} & ReP & Immunofluorescence images & CNN to classify whether there is a relapse of patients with triple-negative breast cancer within 3 years.\\
        
        \cite{yang2022prediction} & ReP & H\&E, NIM & Multimodal Compact Bilinear to fuse features extracted by CNN and clinical features to predict whether there will be a recurrence.\\
        
        \cite{liu2021mask} & ReP & MRI & Modified 3D VGG for 5-year recurrence prediction and HER2 status prediction from dynamic contrast enhanced MRI.\\
        
        \cite{ma2022radiomics} & ReP & MRI  & A 3D U-Net for tumor segmentation + feature extraction from segmented images + SVM for recurrence prediction of triple-negative breast cancer treated with neoadjuvant chemotherapy.\\
        
        \cite{rabinovici2022multimodal} & ReP & MRI, NIM& Ensemble learning framework combining CNN-based models for multi-parametric MRI components classification with clinical history and immunohistochemical markers. \\ 
        
        \cite{kim2021deep} &  ReP  & US, MG, NIM & Weibull Time To Event Recurrent Neural Network taking  clinical characteristics, pathologic characteristics after curative surgery,  laboratory tests, mammography features, and breast sonography features for recurrence time prediction. \\
       
        \cite{nichols2020quantitative} & ReP & MG, pathology images, optical spectral imaging, NIM & Use artifical neural network to predict the OncotypeDx recurrence score with feature extracted from different modalities. \\
        
        \cite{whitney2018quantitative}  & ReP &  H\&E & Use  MLP to classify OncotypeDx recurrence group. \\ 
        
        \cite{bai2021open} & OT & H\&E, NIM &  Cell segmentation + cell classification with neural network for tumor-infiltrating lymphocytes (TIL) assessment in triple-negative breast cancer cohorts. \\
        
        \cite{balkenhol2021optimized} & OT &  H\&E & CNN for registering unmixed multispectral images and corresponding H\&E sections, segmenting the different tissue compartments, and detecting all individual positive lymphocytes.\\
        
        \cite{qiu2022intratumor} & OT  & H\&E, NIM & Graph neural network to discover interactions between biomarkers.\\
        
        \cite{lee2022derivation} & OT &  H\&E& Graph neural network to offer interpretable biomarker. \\
       
        \cite{zuo2022identify} & OT & H\&E, NIM & Graph attention network to describe the spatial interactions of TILs and tumor regions across whole-slide images. \\
        
		\bottomrule
	\end{tabularx}
\end{table*}

\FloatBarrier
\bibliographystyle{sss.bst}
\bibliography{reference}

\end{document}